\documentclass[12pt]{article}
\usepackage{amssymb,amscd,amsmath,amsfonts}
\usepackage{hyperref}

\makeatletter \@addtoreset{equation}{section} \makeatother

\def\abz{\vspace{3mm} \noindent}

\def\dps{\displaystyle}
\newcommand{\half}{\frac{1}{2}}

\def\be{\begin{equation}}
\def\ee{\end{equation}}
\def\ba{\begin{array}}
\def\ea{\end{array}}

\def\ads{AdS_d}
\def\ds{dS_d}
\def\bads{(A)dS_d}

\def\hc{hc(1|2\!:\![d-1,2])}
\def\hs{hu(1|2\!:\![d-1,2])}

\def\d{\partial}

\def\bs{\bar{s}}
\def\bv{\bar{v}}
\def\bta{\bar{\eta}}

\def\cA{\mathcal{A}}

\def\cE{\mathcal{E}}

\def\cH{\mathcal{H}}

\def\cM{\mathcal{M}}

\def\cP{\mathcal{P}}
\def\cQ{\mathcal{Q}}

\def\cS{\mathcal{S}}

\def\dalpha{\dot{\alpha}}

\def\bv{\bar v}

\begin{document}

\renewcommand{\thefootnote}{\fnsymbol{footnote}}


\begin{flushright}
FIAN/TD/22/07
\end{flushright}

\vspace{7mm}

\begin{center}

{\large\textsc{On Manifestly $Sp(2)$ Invariant Formulation
\\
\vspace{1mm}
Of Quadratic Higher Spin Lagrangians}}

\vspace{.9cm}

K. ~B.~ A l k a l a e v
\vspace{.5cm}

\textit{I.E. Tamm Department of Theoretical Physics, \\P.N. Lebedev Physical
Institute,\\ Leninsky ave. 53, 119991 Moscow, Russia\footnote{Permanent address. E-mail: alkalaev@lpi.ru}}

\vspace{2mm}

\textit{and}

\vspace{2mm}

\textit{Scuola Normale Superiore and INFN, \\Piazza dei Cavalieri 7,
I-56126 Pisa, Italy\footnote{Postdoctoral visitor of INTAS and Scuola Normale Superiore.}}

\end{center}

\vspace{.5cm}

\begin{minipage}{.90\textwidth}

\textsc{Abstract.} The Lagrangian frame-like formulation of free higher spin symmetric bosonic
$\ads$ fields is given within a manifestly $sp(2)$ invariant framework. It is designed to deal with
infinite multiplets  of fields appearing as gauge connections of the higher spin algebras.

\end{minipage}

\vspace{.9cm}

\renewcommand{\thefootnote}{\arabic{footnote}}
\setcounter{footnote}{0}

\noindent {\large\textsc{1. Introduction}}
\label{sec:int}
\vspace{5mm}

\abz At the free field level, the dynamics of higher spin (HS) symmetric massless fields is pretty
well understood both on the flat and $\bads$ backgrounds. Various formulations of HS symmetric
field dynamics are available but basically there are two main methods to describe  HS
fields, the metric-like and
the frame-like approaches.
Within the more traditional metric-like formulation
HS fields are described by Lorentz-covariant tensor fields
\cite{fronsdal,Curtright:1979uz,de Wit:1979pe,Siegel:1985tw,Metsaev,Buchbinder:2001bs,Francia:2002aa}
\footnote{The light-cone actions for arbitrary $AdS_5$
mixed-symmetry massless fields were constructed  in \cite{Metsaev:2002vr}.}. The frame-like formulation describes
HS fields as $p$-forms with tangent Lorentz indices of definite symmetry types
\cite{Vasiliev:1980as,V1,Lopatin:1987hz,Vasiliev:2001wa}. Both formulations are dynamically
equivalent and the metric-like fields result from the frame-like ones by virtue of partial
gauge fixing. There exists also the so called parent theory which encodes these two
forms of HS field dynamics and particular realizations can be reached by one or another
reduction \cite{Barnich:2004cr}.

\abz On the $\bads$ backgrounds the HS fields may exhibit an interesting property
of partial masslessness \cite{Deser:1983mm,pmf,Hallowell:2005np,zin,Skvortsov:2006at}. HS fields of this
type possess a reduced gauge symmetry compared to that of massless fields and describe either non-unitary dynamics (for the $\ads$ background) or
dynamics with the energy not bounded from below (for the $\ds$ background). In the flat limit
partially massless fields do not exist and reduce to usual massless fields.

\abz From the group-theoretical point of view, a given formulation of a single free field
should give rise to the corresponding infinite-dimensional representation of the
algebra of global space-time symmetries. More precisely, the space of one-particle states
should form a unitary representation of the Poincare algebra or the $\bads$ algebra with the energy bounded from below.
The conditions
of unitarity or bounded energy  can be relaxed like in the case of partially massless fields.

\abz From the perspective of the higher spin interaction problem, a free field theory is required
to satisfy some additional conditions. The reason is that the higher spin interactions are governed by a
higher spin algebra which describes both global and gauge higher spin symmetries.
It defines a field content of the theory and consistent deformations of linearized
gauge symmetries. This implies that for a free field theory to be a limit of some non-linear theory
the fields must be organized into an infinite higher spin multiplet. In other words,
fields of a given higher spin multiplet form a representation of the algebra of global
higher spin symmetries. This is the so called  admissibility condition \cite{Konshtein:1988yg}.

\abz In this paper we aim to develop a Lagrangian framework for free higher spin dynamics that
naturally operates with infinite sets of  (partially) massless bosonic symmetric fields considered
as gauge connections of the higher spin algebras. Our goal is motivated by a desire to develop a
Lagrangian description of the higher spin couplings in the cubic order and beyond \footnote{The analogous problems
for non-symmetric (partially) massless fields require a separate study. The reason is that higher spin
algebras with spectra of $\ads$ fields with any spins are unknown yet \cite{Vasiliev:2004cm}.
Let us also note that the frame-like formulation for arbitrary non-symmetric massless
fields in $\ads$ is available now \cite{ASV2}.}.

\abz Our considerations rest on recent developments in the non-linear HS theory utilizing
the unfolded form of HS dynamics \cite{Vasiliev:2003ev, Vasiliev:2004cm, Sagnotti:2005ns}.
The main technical ingredient proposed in \cite{Vasiliev:2003ev} is the use of $sp(2)$ symmetry as internal symmetry in the
auxiliary space. More precisely, the unfolded HS dynamics is formulated in terms of
functions $F(x|\,Y)$, which depend on the spacetime coordinates $x^n$ and $sp(2)$ doublets of
the auxiliary $o(d-1,2)$ vector variables $Y^A_\alpha$ and are subject to the $sp(2)$ invariance condition.
It follows that the fields of the theory are identified with the expansion coefficients with
respect to the auxiliary variables.
It is remarkable that the use of the auxiliary $sp(2)$ symmetry  brings together the previously known unfolded field equations and the newly defined HS algebra,
and provides for them a unified framework \footnote{Within the unfolded
formulation one may consider two different versions of the $sp(2)$ invariance condition,
a weak condition originally introduced in \cite{Vasiliev:2003ev}, and a strong one
subsequently proposed in \cite{Sagnotti:2005ns}.
The strong form of $sp(2)$ invariance requires further investigations and
should be treated with a great care when one considers interactions. In the present
paper, we study free gauge fields only and use the $sp(2)$ invariance condition in the form proposed in \cite{Vasiliev:2003ev}.}. Having in mind a Lagrangian form of non-linear dynamics,
these results along with the Lagrangian frame-like formulation  \cite{Lopatin:1987hz,Vasiliev:2001wa}
 provide a good starting point.

\abz Let us recall now the general properties of the HS algebra that describes massless symmetric
bosonic $\ads$ fields of any spins from zero to infinity \cite{Eastwood:2002su,Vasiliev:2003ev}. By
definition, it is a quotient algebra and massless HS fields are identified with the representatives
of the equivalence classes. More precisely, HS algebra is a quotient $S/I$, where $S$ is an
infinite-dimensional Lie algebra endowed with the Weyl $*$-product commutator, which describes
two-row $o(d-1,2)$ traceful tensors, and $I$ is a two-sided ideal generated by traces. It follows
that the $*$-product of any two elements satisfying the tracelessness condition does not
necessarily satisfies the same condition (otherwise these elements would form a subalgebra rather
than a quotient algebra).
It causes the problem of explicit realization of the $*$-product on the factor space. In particular, the structure constants of
the HS algebra are not known yet  in general.
However, there is a
nice projection technique based on the quasi-projector $\Delta$ that allows one to perform a
factorization procedure automatically without an explicit calculation of particular representatives
\cite{Vasiliev:2003ev,Vasiliev:2004cm,Vasiliev:2001wa,Alkalaev:2002rq,Sagnotti:2005ns}.

\abz The algebra $S$ describes infinitely degenerate sets of massless and
partially-massless fields, while the quotient $S/I$ describes massless fields only and each field enters in a single
copy. The factorization procedure removes partially massless fields
and reduces an infinite degeneracy of massless fields. In principle,
a non-linear theory governed  by the algebra $S$ may be of interest. It would provide
an example of the gauge system with an infinitely extended gauge symmetry that contains
conventional massless HS gauge theories. In particular, this point of view leads to
the issue of consistent interactions of partially massless
HS fields with the gravity and between themselves \footnote{It was shown that consistent
cubic vertex of a partially massless spin-5/2 coupled to the $\ads$ gravity does exist \cite{Metsaev:2006ui}.}.
Let us emphasize, that a presence of partially-massless fields
extends gauge symmetry of the theory at the expense of a lack of untarity.
However, it still makes sense to study such extended theories since unitary massless theory
could be embedded  into it by virtue of one or another scenario.

\abz In this paper we propose to implement the  projection technique described above on the level of action
functionals. The procedure has two stages. Firstly, one introduces the action functional $\cS[\Omega]$
defined on the fields $\Omega$ which are elements of the  algebra $S$. Secondly,
one builds the projector $\Delta$ into the action $\cS$ in an appropriate fashion.
The action $\cS_\Delta[\Omega]$ equipped with the projector is formally defined on the elements
of the algebra $S$  but the presence of the projector reduces it to the quotient  $S/I$. In fact,
this approach is inspired by the analysis of  \cite{Vasiliev:2001wa,Alkalaev:2002rq}, where
the similar projection procedure was used in the study of $5d$ HS cubic couplings.

\abz We study action functionals  defined on the fields taking values in the
algebra $S$, while the actions defined on the quotient $S/I$ will be given elsewhere \cite{Alk_next}.
From the technical point of view, when analyzing action functionals we use a particular
framework of \cite{ASV2} developed to describe a frame-like form of  mixed-symmetry field dynamics.
It is remarkable that the approach
of \cite{ASV2} fits
naturally the definition of HS algebra \cite{Vasiliev:2003ev,Vasiliev:2004cm}
thus making the analysis of the problem feasible.

\abz The paper is organized as follows. In section 2 we shortly review frame-like formulation
of symmetric bosonic fields in $d$ dimensions.
Section 3 contains a discussion of $sp(2)$ doublets of $o(d-1,2)$ vector variables and
the trace decomposition of rectangular traceful $o(d-1,2)$ tensors. In section 4 we introduce the
bilinear form defined on arbitrary $sp(2)$ singlet fields, which serves as a basis for the action
functionals discussed in section 5. In particular, in section 5.3 we build HS action
that describes an infinite set of massless gauge fields with spins $2\leq s\leq \infty$ and reproduces
known expression for the component form of HS action elaborated in Ref. \cite{Vasiliev:2001wa}.
Finally, we present our conclusions in section 6. In Appendix we
collect and discuss some useful formulae used in the calculations.

\vspace{10mm}
\noindent {\large\textsc{2. A  summary on HS symmetric fields}}
\label{sec:alg}
\vspace{5mm}

\noindent Let us briefly recall some basic facts on the frame-like formulation
of HS symmetric fields. More detailed
expositions can be found, for instance, in \cite{obzor} and \cite{Skvortsov:2006at}.

\abz Within the metric-like approach an integer spin-$s$ massless field is described by a
totally symmetric rank-$s$ $o(d-1,1)$ tensor
$\varphi^{a_1\ldots\, a_s}$
subject to the Fronsdal double tracelessness condition \cite{fronsdal}.
The  higher spin  gauge transformations read
$\delta\varphi^{a_1\ldots\, a_s}={\cal D}^{(a_1} \varepsilon^{a_2\ldots\, a_s)}\,,$
where the parameter $\varepsilon^{a_1\ldots\, a_{s-1}}$ is a rank ($s-1$) symmetric
traceless tensor and ${\cal D}^a$ is
the background Lorentz derivative \footnote{We work within the mostly
minus signature and use notation $\underline{m},\underline{n} =
0\div d-1\;$ for world indices, $a,b= 0\div d-1$ for tangent
Lorentz $o(d-1,1)$ vector indices and $A,B = 0 \div d $ for
tangent $\ads$ $o(d-1,2)$ vector indices. We also use the condensed
notation of \cite{V1} and denote a set of symmetric indices $(a_1 \cdots a_s)$
as $a(s)$. All symmetrizations are performed with
a unit weight, \textit{e.g.} $X^{(a}Y^{b)}=X^a Y^b+X^b Y^a$. }.

\abz The frame-like formalism  operates with 1-form field \cite{Vasiliev:2001wa}
\be
\label{frO} \Omega^{A_1\ldots\, A_{s-1},\, B_1\ldots\, B_{s-1}} =\textrm{d}x^{\underline{n}}\,
\Omega_{\underline{n}}{}^{A_1\ldots\, A_{s-1},\, B_1\ldots\, B_{s-1}}
\ee
that carries the traceless tensor representation of  $o(d-1,2)$ described by the length $s-1$
two-row rectangular Young tableau, {\it i.e.},  symmetrization of any $s$ tangent indices of
(\ref{frO}) gives zero.

\abz With the $1$-form field (\ref{frO}) one associates the linearized curvature
\be
\label{R1A} R^{A_1\ldots\, A_{s-1},\, B_1\ldots\, B_{s-1}} = D_0
\Omega^{A_1\ldots\, A_{s-1},\, B_1\ldots\, B_{s-1}}\,,
\ee
where $D_0T^A={\rm d}T^A+\Omega_0^A{}_BT^B$ is the $o(d-1,2)$ covariant derivative
evaluated with respect to the $\ads$ background $1$-form connection
$\Omega_0^{AB}=-\Omega_0^{BA}$ that satisfies the zero curvature equation
$D_0D_0={\rm d}\Omega_0^{AB}+\Omega_0^A{}_C\wedge\Omega_0^C{}_B=0$. The last property implies
that the curvature (\ref{R1A}) is  invariant under  HS gauge
transformations
\be
\label{litr}
\delta \Omega^{A_1\ldots\, A_{s-1},\, B_1\ldots\, B_{s-1}}
= D_0 \varepsilon^{A_1\ldots\, A_{s-1},\, B_1\ldots\, B_{s-1}}
\ee
with a traceless 0-form gauge parameter.

\abz Being decomposed into $o(d-1,1)$ components,  the  field (\ref{frO}) yields
a collection of  $1$-forms \cite{Lopatin:1987hz}
\be
\label{sym_fr_fields}
\omega^{a_1\ldots\, a_{s-1},\, b_1\ldots\, b_t}(x) =
\textrm{d}x^{\underline{n}}\, \omega_{\underline{n}}{}^{a_1\ldots\, a_{s-1},\, b_1\ldots\, b_t}(x)\;,
\qquad
0\leq t \leq s-1\;.
\ee
The 1-form  with $t=0$ is the physical field $\omega_{\underline{n}}{}^{a(s-1)}$.
The 1-form with $t=1$ is the auxiliary field $\omega_{\underline{n}}{}^{a(s-1),\,b}$.
The remaining 1-forms (\ref{sym_fr_fields}) with $t\geq 2$ are extra fields
\cite{V1,Lopatin:1987hz}. The metric-like field is identified with a component of
the physical field obtained by the symmetrization $\omega^{(a_1;\,a_2 ... a_s)}(x)= \varphi^{a_1 ... a_s}(x)$.

\abz Lorentz fields play different dynamical roles depending on the values of parameter $t$.
For example, in  the  spin two case the
decomposition has the form $\Omega^{AB}\rightarrow \omega^a \oplus
\omega^{ab}$, where $\omega^a$ is the vielbein  (the physical
field) and $\omega^{ab}=-\omega^{ba}$ is the Lorentz spin
connection (the auxiliary field). On the level of the equations of motion the auxiliary
field is expressed through the first derivatives of the physical field.
Extra fields are absent in this case. They appear starting from a spin three field and
are required to enter the action via total derivatives only.
Their role is to maintain a manifest gauge invariance of the action built as a bilinear
combination of linearized curvatures (\ref{R1A}) (for more details, see Ref. \cite{obzor} and
sections  4, 5 of the present paper).

\abz The decomposition procedure of $o(d-1,2)$ covariant fields into their $o(d-1,1)$
irreducible components can be done in a manifestly $o(d-1,2)$ covariant fashion \cite{compensator}.
To this end one identifies
the Lorentz algebra as a stability subalgebra of the compensator vector $V^A$ normalized as $V^AV_A=1$.
Then the resulting set of Lorentz components can be described as $o(d-1,2)$ tensors which are
orthogonal to the compensator vector.

\abz In particular, the component of $\Omega^{A(s-1),\, B(s-1)}$, that is most parallel to the
compensator $V^A$, is the physical field
$\omega^{A(s-1)} = \Omega^{A(s-1), B(s-1)} V_B \cdots V_B$. (In order to obtain manifestly Lorentz covariant expressions
it is convenient to substitute the compensator in the form $V^A = \delta^A{}_{d}$.)
The less $V^A$-longitudinal components are identified with the other fields in the
set (\ref{sym_fr_fields}).

\abz It is useful to introduce a field
\be
\label{phaux}
\hat \Omega^{A_1\ldots\, A_{s-1},\, B_1\ldots  B_p}
\equiv
\Omega^{A_1\ldots\, A_{s-1},\, B_1\ldots  B_p C_{p+1}\ldots\, C_{s-1}}\; V_{C_{p+1}} \cdots V_{C_{s-1}}\;,
\quad
p\leq s-1\;,
\ee
defined as a contraction of the original field (\ref{frO}) with a number of compensators \footnote{Let us
note that by virtue of Young symmetry properties a contraction with $s$ compensators gives zero. Also,
any contraction of a rectangular Young tableau
with the   symmetric tensor $V_{A_1 \ldots A_k}=V_{A_1}\cdots V_{A_k}$ can be reduced to a contraction
of $V_{A_1 \ldots A_k}$ with indices of the bottom row.}. It
decomposes into a set of  Lorentz fields (\ref{sym_fr_fields}) with $0 \leq t \leq p\;$.
In the case of $p=1$ the field $\hat \Omega^{A(s-1),\, B}$ contains just two Lorentz
components which are identified with the physical ($t=0$) and the auxiliary ($t=1$) fields.

\abz The $o(d-1,2)$ covariant versions of the background frame  and background Lorentz
spin connection are defined as follows \cite{compensator}
\be \label{s2_def_lor}
\lambda E_0^A = D_0\,V^A\equiv {\rm d} V^A+\Omega_0^{AB}V_B\;, \quad
\omega_0^{AB} = \Omega_0^{AB} -\lambda\,(E_0^{A}\,V^{B}-E_0^{B}\,V^{A})\;.
\ee
Here the parameter $\lambda$ is the inverse radius of the $\ads$ space.

\abz The frame-like formulation of HS dynamics provides
a description for a  wider class of relativistic fields that may propagate on the $\bads$
backgrounds known as partially massless fields \cite{pmf}.
Indeed, according to Ref. \cite{Skvortsov:2006at} a partially-massless field of spin $s$ and
depth $t$ can be described by $1$-form
\be
\label{frOPM} \Omega^{A_1\ldots\, A_{s-1},\, B_1\ldots\, B_{t}} =\textrm{d}x^{\underline{n}}\,
\Omega_{\underline{n}}{}^{A_1\ldots\, A_{s-1},\, B_1\ldots\, B_{t}}\;,
\ee
that carries the traceless tensor representation of  $o(d-1,2)$ described by two-row
non-symmetric Young tableau with lengths of rows $s-1$ and $t$. The corresponding
field-theoretical systems describe either non-unitary dynamics (for the $\ads$ background) or
dynamics with the energy not bounded from below (for the $\ds$ background).

\abz The linearized curvatures and the gauge transformations are defined in the same manner as for massless
fields
\be
\label{pcurv}
R^{A_1\ldots\, A_{s-1},\, B_1\ldots\, B_{t}} = D_0
\Omega^{A_1\ldots\, A_{s-1},\, B_1\ldots\, B_{t}}\,,
\ee
\be
\label{pgauge}
\delta \Omega^{A_1\ldots\, A_{s-1},\, B_1\ldots\, B_{t}}
= D_0 \varepsilon^{A_1\ldots\, A_{s-1},\, B_1\ldots\, B_{t}}\;,
\ee
and  reproduce the expressions (\ref{R1A}) and (\ref{litr}) at $t=s-1$.

\abz Gauge transformations for partially-massless fields written in a metric-like form involve higher derivatives
up to $(s-t)$-th order \cite{pmf}. It is worth to note that gauge transformations for frame-like fields,
both massless and partially-massless, involve just one derivative, as is seen from (\ref{pgauge}).

\abz In what follows we assume that the parameter $t$ is free and runs from $0$ to $s-1$
so both massless and partially-massless fields are considered on equal footing.

\vspace{10mm}
\noindent {\large\textsc{3. $\;\;$ Non-symmetric tensors and Howe duality}}
\label{sec:actfock}
\vspace{5mm}

\noindent In this section we discuss auxiliary $o(d-1,2)$ vector variables and their Howe dual
$sp(2)$ algebra. Although much of the discussion is familiar from Refs.
\cite{Vasiliev:2003ev,Vasiliev:2004cm}, we feel it is important to set notation carefully, as we
will introduce some new ingredients.

\vspace{10mm}
\noindent \textit{3.1. Auxiliary $sp(2)$ covariant variables}
\label{sec:actfock}
\vspace{5mm}

\noindent Let $Y^A_{\alpha}$ be auxiliary commuting variables \footnote{In the sequel we introduce additional
set of auxiliary dotted variables $Y^A_{\dot \alpha}$, $\dot \alpha=1,2$. All constructions of this section
for undotted variables $Y^A_{\alpha}$ are valid for dotted ones as well.} with $A=0\div d$ and $\alpha = 1,2$. Indices $A$ are
raised and lowered with the invariant symmetric form  $\eta_{AB}$ of $o(d-1,2)$ as
$X_A=\eta_{AB}X^B$, indices $\alpha$ will  be later on specified to $sp(2)$.

\abz The expansion coefficients of polynomial of given orders $m_1$, $m_2$ in variables $Y_1^A$ and $Y^B_2$
\be
\label{dec_two}
F(Y) = F_{A_1...A_{m_1};\; B_1 ... B_{m_2}}\; Y_{1}^{A_1} \cdots
Y_{1}^{A_{m_1}}\;Y_{2}^{B_1} \cdots Y_{2}^{B_{m_2}}
\ee
are $o(d-1,2)$ tensors that are symmetric in the indices $A_i$ and $B_j$. To specify a particular
Young symmetry type of indices one introduces operators
\be
L_\alpha{}^\beta= Y_\alpha^A\,\frac{\d}{\d Y_\beta^A}\;
\ee
that form the $gl(2)$ algebra
$[L_\alpha{}^\beta, L_\gamma{}^\rho] = \delta_\alpha{}^\rho L_\gamma{}^\beta - \delta_\gamma{}^\beta L_\alpha{}^\rho$, which
is Howe dual symmetry \cite{howe,Vasiliev:2003ev, Vasiliev:2004cm}.
Young symmetry conditions read then
\be
\label{Y1}
L_\alpha{}^\beta\, F(Y)\Big|_{\alpha<\beta} = 0\;,
\qquad
L_\alpha{}^\beta\, F(Y)\Big|_{\alpha=\beta} = m_\alpha \,F(Y)\;,
\ee
and mean that $F(Y)$ belongs to a highest weight representation of $gl(2)$ \cite{howe,Vasiliev:2003ev, Vasiliev:2004cm}.
It results in a standard symmetrization condition for the expansion coefficients
\be
\label{Young}
F^{(A_1 \ldots A_{m_1};A_{m_1+1})B_2\ldots B_{m_2}} = 0\;.
\ee

\vspace{3mm}

\noindent An important observation  is
that Young symmetry conditions specifying a block tableau ($m_1=m_2$) can be reformulated as an
invariance of $F(Y)$ under $sl(2)\subset gl(2)$ transformations \cite{howe,Vasiliev:2003ev, Vasiliev:2004cm}. By definition, $sl(2)$ generators
are traceless parts of $gl(2)$ ones
\be
\label{slgen}
\tilde L_\alpha{}^\beta = L_\alpha{}^\beta - \half \delta_\alpha{}^\beta N\;,
\qquad
[\tilde L_\alpha{}^\beta, \tilde L_\gamma{}^\rho] = \delta_\alpha{}^\rho \tilde L_\gamma{}^\beta
- \delta_\gamma{}^\beta \tilde L_\alpha{}^\rho\;,
\ee
(here $N=L_\gamma{}^\gamma$ is a central element of $gl(2)$), so the $sl(2)$ invariance is expressed by
the condition
\be
\label{sl}
\tilde L_\alpha{}^\beta F(Y) = 0\;.
\ee
In other words, polynomial $F(Y)$ with coefficients being two-row rectangular tableaux is
a $sl(2)$ singlet. Note that $N$ plays the role of the Euler operator that counts a total number of
the variables $Y_\alpha^A$.

\abz Now we recall the well-known fact that $2d$ Levi-Civita symbol $\epsilon_{\alpha\beta}$ is an invariant tensor
of $sl(2)$ algebra. In particular, it makes the isomorphism $sl(2)\sim sp(2)$ clear.
From now on we assume that indices $\alpha$ are
raised and lowered with $sp(2)$ invariant antisymmetric form $\epsilon_{\alpha\beta}$  as
$X^\alpha=\epsilon^{\alpha\beta}X_\beta$, $X_\alpha=X^\beta \epsilon_{\beta\alpha}$.

\abz The generators of $sp(2)$ expressed via $sl(2)$ ones
\be
\label{spgen}
T_{\alpha\beta} = \epsilon_{\alpha\gamma}\tilde L_\beta{}^\gamma + \epsilon_{\beta\gamma}\tilde L_\alpha{}^\gamma
\equiv \epsilon_{\alpha\gamma}L_\beta{}^\gamma + \epsilon_{\beta\gamma}L_\alpha{}^\gamma
\ee
form the $sp(2)$ algebra
$
[T_{\alpha\beta}, T_{\gamma\rho}] = -\epsilon_{\alpha\gamma} T_{\beta\rho} - \epsilon_{\alpha\rho} T_{\beta\gamma}
-\epsilon_{\beta\gamma} T_{\alpha\rho}-\epsilon_{\beta\rho} T_{\alpha\gamma}\;.
$
Variables $Y^A_\alpha$ rotate as $sp(2)$ vectors
\be
[T_{\alpha\beta}, Y_\gamma^A] = \epsilon_{\alpha\gamma} Y_\beta^A+\epsilon_{\beta\gamma} Y_\alpha^A\;.
\ee
An equivalent form of (\ref{sl}) reads now
\be
\label{sl_eq}
T_{\alpha\beta}F(Y) \equiv \Big(\epsilon_{\alpha\gamma}L_\beta{}^\gamma + \epsilon_{\beta\gamma}L_\alpha{}^\gamma\Big)F(Y)=0\;.
\ee
The analog of formula (\ref{slgen})
\be
T_{\alpha\beta} = 2 L_\alpha{}^\gamma\epsilon_{\gamma\beta}  - \epsilon_{\alpha\beta} N
\ee
is useful in calculations and is a simple consequence of rank-2 tensor decomposition
into (anti)symmetric parts.

\vspace{7mm}
\noindent \textit{3.2. Trace decomposition}
\label{sec:actfock}
\vspace{5mm}

\noindent Consider a polynomial $F(Y)$ subject to $sp(2)$ invariance condition (\ref{sl_eq}).
The expansion coefficients of $F(Y)$ are two-row rectangular Young tableaux $F^{A_1 \ldots\, A_{m};\,B_1\ldots\, B_{m}}\equiv F^{A(m),\, B(m)}$.
In general, tensors $F^{A(m),\, B(m)}$ are traceful.

\abz  To examine a decomposition of a traceful tensor $F^{A(m), B(m)}$ into traceless components
one needs to study the symmetry properties of $n$-valued product of $o(d-1,2)$ invariant tensors
$\eta_{AB}$. Coupling $n$ traces to  $F^{A(m),\,
B(m)}$   means that one takes a symmetrized tensor product of $n$ traces and then projects out
the components with more than two rows. Graphically, this operation is represented as follows

\vspace{2mm}

\be
\label{tracedecgraph}
\begin{picture}(212,05)(-1,4)
\unitlength=0.6mm
\linethickness{0.210mm}
\put(-10,05){$\cP\Big($}
\put(00,10){\line(1,0){10}} 
\put(00,05){\line(1,0){10}} 
\put(00,05){\line(0,1){5}} 
\put(05,05){\line(0,1){5}} 
\put(10,05){\line(0,1){5}}

\put(15,05){$\otimes\; \cdots \;\otimes$}

\put(45,10){\line(1,0){10}} 
\put(45,05){\line(1,0){10}} 
\put(45,05){\line(0,1){5}} 
\put(50,05){\line(0,1){5}} 
\put(55,05){\line(0,1){5}}

\put(17,-2){\footnotesize $n \;\; times$}

\put(56,05){$\dps\Big)_{\rm sym}=\bigoplus_{2l+k=n}^{}$}

\put(95,12){\line(1,0){40}} 
\put(95,2){\line(1,0){20}} 
\put(95,7){\line(1,0){40}} 
\put(110,2){\line(0,1){10}} 
\put(105,2){\line(0,1){10}} 
\put(100,2){\line(0,1){10}} 
\put(95,2){\line(0,1){10}} 
\put(115,2){\line(0,1){10}} 

\put(135,7){\line(0,1){5}} 
\put(130,7){\line(0,1){5}} 
\put(125,7){\line(0,1){5}} 
\put(120,7){\line(0,1){5}} 
\put(115,7){\line(0,1){5}} 

\put(107,-5){\footnotesize $2l$}
\put(107,15){\footnotesize $2l+2k$}
\put(150,0){$,$}
\end{picture}
\ee

\vspace{5mm}
\noindent where $\cP$ is a projector on the two-row tensors arising in the tensor product. Obviously,
one can take any number of non-trivial traces from $0$ to $2\,[m/2]$. It follows that
for an arbitrary two-row rectangular traceful tensor there is a two-parametric family of components
\be
\label{tracedec}
F^{A(m),\,B(m)} = \bigoplus_{l=0}^{[m/2]}\,\bigoplus_{k=0}^{[m/2]-l}\; F^{A(m-2l),\,B(m-2l-2k)}\;.
\ee
Here the parameter $2l+k=n$ counts a number of removed traces and two-row tensors in the
right-hand-side are traceless
\be
\eta_{A(2)}F^{A(p),\,B(t)}=0\;,
\quad
\eta_{AB}F^{A(p),\,B(t)}=0\;,
\quad
\eta_{B(2)}F^{A(p),\,B(t)}=0\;.
\ee
Note that by virtue of Young symmetry (\ref{Young}) only the first condition is independent while the others
are its linear combinations. For a block tableau, \textit{i.e.} when  $p=t$, the first and the last
conditions are equivalent because of a block symmetry property $F^{A(m),\,B(m)} = (-)^m F^{B(m),\,A(m)}$.

\abz We now turn to a reformulation of the above results within a manifestly $sp(2)$ covariant framework. It heavily
rests on a possibility to describe non-symmetric two-row tensors as $sp(2)$ singlets. The only
restriction is that a difference between the lengths of first and second rows must be even.

\abz Let us introduce  the operators
\be
\label{tropers}
t_{\alpha\beta} = \eta_{AB} \,Y_\alpha^AY_\beta^B\;
\quad
{\rm and}
\quad
\bs^{\alpha\beta} = \eta^{AB} \,\frac{\d^2}{\d Y_\alpha^A \d Y_\beta^B}\;.
\ee
Their commutation relation is given by
\be
\label{comrel}
\ba{l}
\dps[\bs^{\alpha\beta}, t_{\gamma\rho}] =
(\frac{N}{2}+d+1)(\delta_\gamma{}^\beta\delta_\rho{}^\alpha+\delta_\rho{}^\beta\delta_\gamma{}^\alpha)+
\\
\\
\dps
\qquad\qquad\qquad\qquad+
\half(\delta_\gamma{}^\beta T_\rho{}^\alpha+\delta_\rho{}^\beta T_\gamma{}^\alpha+
\delta_\gamma{}^\alpha T_\rho{}^\beta+\delta_\rho{}^\alpha T_\gamma{}^\beta)\;.
\ea
\ee
Note also that the operators $t_{\alpha\beta}$, $\bs^{\gamma\rho}$ and
$\dps l_\mu{}^\nu=L_\mu{}^\nu + \frac{d+1}{2}\,\delta_\mu{}^\nu$
form the $sp(4)$ algebra, which is the Howe dual algebra for traceless two-row
tensors \cite{howe,Vasiliev:2003ev, Vasiliev:2004cm} \footnote{
The space of polynomials $F(Y)$ carries representations of Howe dual algebras,
$o(d-1,2)$ and $sp(4)$ (and $sp(2)\subset sp(4)$). Since these algebras commute the space $F(Y)$ decomposes into a direct
sum of irreducible highest weight $sp(4)$ representations. A non-trivial statement proved in
\cite{Barnich:2004cr} is that for $d\geq3$ each  irreducible $sp(4)$ component  is an infinite-dimensional
generalized Verma module induced from a finite-dimensional $sp(2)$ representation  and does not
not contain singular vectors except the trivial one generated by the $sp(2)$ generator.
I am grateful to M. Grigoriev for the illuminating discussions of this
issue.}.

\abz Since traceless $o(d-1,2)$ tensors  satisfy
the constraint
\be
\label{traceless}
\bs^{\alpha\beta} F(Y) = 0\;,
\ee
it follows that
$t_{\alpha\beta}$ and $\bs^{\gamma\rho}$  act like a trace creation and a trace annihilation
operators. They are symmetric tensors with respect to the $sp(2)$ transformations
\be
\ba{l}
[T_{\alpha\beta}, t_{\gamma\rho}]
= \epsilon_{\alpha\gamma} t_{\beta\rho} +\epsilon_{\beta\gamma} t_{\alpha\rho}+
\epsilon_{\alpha\rho} t_{\beta\gamma}+\epsilon_{\beta\rho} t_{\alpha\gamma}\;,
\\
\\

[T_{\alpha\beta}, \bs^{\gamma\rho}] =
\delta_\beta{}^\rho \bs_\alpha{}^\gamma + \delta_\alpha{}^\rho \bs_\beta{}^\gamma
+\delta_\beta{}^\gamma \bs_\alpha{}^\rho+\delta_\alpha{}^\gamma \bs_\beta{}^\rho\;.

\ea
\ee

\abz Consider now $o(d-1,2)$ tensors that are not traceless, \textit{i.e.},
\be
\label{traceful}
\bs^{\alpha\beta} F(Y) \neq 0.
\ee
Taking into account the commutation relation (\ref{comrel}) one finds that the general solution
of the condition (\ref{traceful}) reads
\be
\label{trdec}
F(Y) = F_0(Y)+t_{\alpha\beta}\, F_1^{\alpha\beta}(Y)\;,
\ee
where $F_0(Y)$  satisfies (\ref{traceless}), while  $F_1^{\alpha\beta}(Y)$ is
a symmetric $sp(2)$ tensor that rotates as
\be
\label{sp_trans}
[T^{\alpha\beta}, F_1^{\gamma\rho}] =
\epsilon^{\alpha\gamma} F_1^{\beta\rho} + \epsilon^{\beta\gamma} F_1^{\alpha\rho}
+\epsilon^{\alpha\rho} F_1^{\beta\gamma}+\epsilon^{\beta\rho} F_1^{\alpha\gamma}\;,
\ee
thus a combination $t_{\alpha\beta}F_1^{\alpha\beta}$ remains invariant under the $sp(2)$
transformations \cite{Vasiliev:2003ev}.

\abz Let us suppose that $\bs^{\alpha\beta} F_1^{\gamma\rho}(Y)=0$. Applying the trace annihilation
operator $\bs^{\alpha\beta}$ to both sides of the relation (\ref{trdec}) we obtain that
\be
\bs^{\alpha\beta} F(Y) = (N+2d+10)\, F_1^{\alpha\beta}(Y)\;,
\qquad
\bs^{\alpha\beta} \bs^{\gamma\rho} F(Y) =0\;.
\ee
These expressions imply that $F(Y)$ describes a double traceless $o(d-1,2)$ tensor. A peculiarity caused by a
manifest $sp(2)$ covariance of the whole analysis consists in the illusory mismatch
between a number of traceless components and a number of trace annihilation operators (\ref{tropers}).
Indeed, according to the trace decomposition (\ref{tracedec}), a double traceless two-row rectangular tensor $F^{A(m),\, B(m)}$ decomposes into
traceless components as follows
\be
\label{decgraph}
F^{A(m),\, B(m)} = F_0^{A(m),\, B(m)} \oplus F_1^{A(m),\, B(m-2)}\;.
\ee
By comparing to (\ref{trdec}) one identifies the first term in (\ref{decgraph}) with $F_{0}(Y)$ and
the second one with $F_1^{\alpha\beta}(Y)$. Then one observes  that the non-symmetric tensor
corresponding to the first trace of $F$ is described by three functions $F_1^{11}, F_1^{12}$ and
$F_1^{22}$ and not by a single one as one might expect. However, commutation relation
(\ref{sp_trans}) ensures that not all functions $F_1^{\alpha\beta}$ are independent and can be
expressed in terms of just one, for instance, $F_1^{22}$. More precisely, particular relations
$[T^{11}, F^{22}] = 4F^{12}$ and $[T^{11}, F_1^{12}] = 2F_1^{11}$ read off from (\ref{sp_trans})
imply that one rotates $F^{22}$ to obtain $F^{11}$ and $F^{12}$. The expansion coefficients
of function $F^{22}$ are $F_1^{A(m),\,B(m-2)}$, while $F^{12}, F^{11}$ describe $F_1^{A(m-1)B,\,B(m-2)}$ and
$F_1^{A(m-2)B(2),\,B(m-2)}$, respectively. The action of generator $T^{11}=-2 L_2{}^1$ is in fact a
proper symmetrization that lengthens a second row.

\abz The $sp(2)$ covariant description of more than double traceless tensors considered above is quite analogous.
A general decomposition of order $2m$ polynomial  $F(Y)$ to traceless parts reads
\be
\label{gendec}
F(Y) = \sum_{n=0}^{2 [m/2]} t_{\alpha_1\beta_1} \cdots t_{\alpha_n\beta_n}\,
F_n^{\alpha_1\beta_1;\;\cdots ;\; \alpha_n\beta_n}(Y)\;,
\quad
\bs^{\gamma\rho}\,F_n^{\alpha_1\beta_1;\;\cdots ;\; \alpha_n\beta_n}(Y) = 0\;.
\ee
The $sp(2)$ representation carried by $F_n^{\alpha_1\beta_1;\;\cdots ;\; \alpha_n\beta_n}(Y)$
is described by a symmetric tensor product of $n$ pairs of symmetrized $sp(2)$ indices. Evaluating
the tensor product  yields the following decomposition

\vspace{2mm}

\be
\label{gendecgr}
\begin{picture}(212,05)(-1,4)
\unitlength=0.6mm
\linethickness{0.210mm}
\put(-10,05){$\Big($}
\put(00,10){\line(1,0){10}} 
\put(00,05){\line(1,0){10}} 
\put(00,05){\line(0,1){5}} 
\put(05,05){\line(0,1){5}} 
\put(10,05){\line(0,1){5}}

\put(15,05){$\otimes\; \cdots \;\otimes$}

\put(45,10){\line(1,0){10}} 
\put(45,05){\line(1,0){10}} 
\put(45,05){\line(0,1){5}} 
\put(50,05){\line(0,1){5}} 
\put(55,05){\line(0,1){5}}

\put(17,-2){\footnotesize $n \;\; times$}

\put(56,05){$\dps\Big)_{\rm sym}=\bigoplus_{2l+k=n}$}

\put(95,12){\line(1,0){40}} 
\put(95,2){\line(1,0){20}} 
\put(95,7){\line(1,0){40}} 
\put(110,2){\line(0,1){10}} 
\put(105,2){\line(0,1){10}} 
\put(100,2){\line(0,1){10}} 
\put(95,2){\line(0,1){10}} 
\put(115,2){\line(0,1){10}} 

\put(135,7){\line(0,1){5}} 
\put(130,7){\line(0,1){5}} 
\put(125,7){\line(0,1){5}} 
\put(120,7){\line(0,1){5}} 
\put(115,7){\line(0,1){5}} 

\put(107,-5){\footnotesize $2l$}
\put(107,15){\footnotesize $2l+2k$}
\put(150,0){$,$}
\end{picture}
\ee

\vspace{5mm}
\noindent which, in fact, gives a collection of $sp(2)$ tensors of the same symmetry types
as the decomposition (\ref{tracedecgraph}) for $o(d-1,2)$ tensors. The only  difference is that
no projection on two-row tensors is needed because in two dimensions  any tensor with more than two rows is identically equal
to zero. Moreover, any tensor with two antisymmetric indices $X^{\alpha\beta}$ can be dualized to a scalar
by virtue of the Levi-Civita tensor $\epsilon_{\alpha\beta}$ as
$\dps X^{\alpha\beta}=\half\epsilon^{\alpha\beta}\,X$, where $X=\epsilon_{\gamma\rho}X^{\gamma\rho}$.
Upon application of these rules,  the decomposition (\ref{gendec}) can be cast into the following form
\be
\label{decc}
F(Y) = \sum_{l=0}^{[m/2]}\sum^{[m/2]-l}_{k=0}\;t_{\alpha_1\ldots\, \alpha_{2k}}\, Z_+^{l}\;
F_{2l+k}^{\alpha_1\ldots \,\alpha_{2k}}(Y)\;,
\quad
\bs^{\gamma\rho} F_{2l+k}^{\alpha_1\ldots \,\alpha_{2k}}(Y)=0\;,
\ee
where the notation are introduced
\be
\label{deccc}
t_{\alpha_1\ldots \alpha_{2k}} = t_{(\alpha_1\alpha_2}\cdots t_{\alpha_{2k-1}\alpha_{2k})}
\;\;\; {\rm and} \;\;\;
Z_+=t_{\alpha\beta}t^{\alpha\beta}\;.
\ee
The quantities $t_{\alpha(2k)}$ and $F_{2l+k}^{\alpha(2k)}(Y)$ are
rank-$2k$ symmetric  $sp(2)$ tensors, and $Z_+$ is $sp(2)$ invariant.
The expansion coefficients of  $F_{2l+k}^{\alpha(2k)}(Y)$ are traceless tensors
$F^{A(m-2l),\, B(m-2l-2k)}$ with the difference between lengths of first and second rows equal to
$2k$. The expressions (\ref{decc}) and (\ref{deccc}) provide $sp(2)$ covariant reformulation of
the trace decomposition (\ref{tracedec}).

\vspace{3mm}

\noindent The above decomposition can be represented in a more compact form that will be used in the sequel. To this end let us slightly
change the notation and introduce
\be
\label{newF}
F_{p,\,t}(Y) = t_{\alpha_1\ldots\, \alpha_{2t}} F_{p,\,t}^{\alpha_1\ldots\, \alpha_{2t}}(Y)\;,
\qquad
\bs^{\gamma\rho} F_{p,\,t}^{\alpha_1\ldots \,\alpha_{2t}}(Y)=0\;,
\ee
where $p\geq 2t$ and $F_{p,\,t}^{\alpha(2t)}(Y)$ describe traceless $o(d-1,2)$
tensors $F^{A(p),\,B(p-2t)}$. Functions $F_{p,\,t}(Y)$ are $sp(2)$ invariant
\be
T_{\alpha\beta}\, F_{p,\,t}(Y)=0\;,
\ee
and subject to a generalized traceless condition
\be
(\bs^{\alpha\beta})^{t+1} F_{p,\,t}(Y) = 0\;.
\ee
The above consideration allows one to say that modulo trace contributions the quantity
$F_{p,\,t}(Y)$ provides an $sp(2)$ invariant description of non-symmetric two-row
$o(d-1,2)$ traceless tensors \footnote{More precisely, the function $F_{p,\,t}(Y)$ provides an $sp(2)$ invariant
description of non-symmetric $o(d-1,2)$ tensors with a minimal number of traces involved.}. For $t=0$ one reproduces the case of traceless rectangular
$o(d-1,2)$ tensors (\ref{traceless}).

\vspace{3mm}

\noindent With the help of functions (\ref{newF}) the trace decomposition becomes now the manifestly $sp(2)$
invariant
\be
\label{decspinv}
F(Y) = \sum_{k=0}^{\textstyle[\frac{m}{2}]} \,\sum^{{\textstyle[\frac{m}{2}]}-k}_{l=0}\, Z_+^{l}\;
F_{m-2l,\,k}(Y)\;.
\ee
In this form it admits a direct generalization to the case when an $sp(2)$ singlet $F(Y)$ is an infinite
power series in the auxiliary variables, \textit{i.e.},
\be
\label{denot}
F(Y)=\sum_{m=1}^\infty F^{(m)}(Y)\;,
\ee
where $F^{(m)}$ is a polynomial of $2m-2$ order in variables $Y^A_\alpha$,
$F^{(m)}(tY)=t^{2m-2}F^{(m)}(Y)$.
The functions $F^{(m)}(Y)$ are $sp(2)$ singlets and  their
expansion coefficients are two-row rectangular traceful $o(d-1,2)$ tensors. By making appropriate field redefinitions
and resummations, a trace decomposition for (\ref{denot}) that generalizes (\ref{decspinv})
can be cast into the form
\be
\label{infexp}
F(Y) = \sum_{n=0}^{\infty}\sum_{k=0}^{\infty}\sum_{m=0}^{\infty}\;
\rho(k,m,n) \,Z_+^n \,F_{k, m;\,n}(Y)\;,
\ee
where $\rho(k,m,n)$ are some non-zero normalization coefficients and the label $n$ in
$F_{k, m;\,n}(Y)$ is introduced to mark an $n$-th copy of $F_{k, m}(Y)$ (\ref{newF}).
It follows that traceless tensors of various symmetry types $F^{A(k),\,B(k-2m)}$ originated
from $F(Y)$ (\ref{denot}) are arranged into an infinite sequence enumerated by a degree of
the quantity $Z_+$.

\vspace{10mm}
\noindent {\large\textsc{4. Bilinear form}}
\label{sec:act}
\vspace{5mm}

\noindent As a preamble to the following we recall that within the frame-like formulation the
action functional for a HS bosonic $\ads$ field
(massless and partially massless) has the following schematic
form
\be
\label{action0}
\cS_2[\Omega] = \int_{\cM^d}\; H^{\textstyle\cdots}(V)\;\epsilon^{\textstyle\cdots}{}_{M_1{\textstyle\ldots} M_{d-4}N}
E_0^{M_1}\wedge\cdots
\wedge E_0^{M_{d-4}}V^N\wedge\; R^{\textstyle\cdots} \wedge
\,R^{\textstyle\cdots}\,,
\ee
where $H^{\textstyle\cdots} (V)$ are some $o(d-1,2)$ covariant coefficients which parameterize various types of index contractions
between curvatures, compensators and the $(d+1)$-dimensional Levi-Civita symbol. Any such action is manifestly $o(d-1,2)$
covariant and gauge invariant with respect to the gauge transformations (\ref{litr}), (\ref{pgauge}). For this action
to describe the correct HS field dynamics, a function $H^{\textstyle\cdots}(V)$ should be fixed by the extra field decoupling condition
\cite{V1,Lopatin:1987hz, Vasiliev:2001wa,Skvortsov:2006at}.

\abz The frame-like action (\ref{action0}) fixed by the extra field decoupling condition
can be reduced to the metric-like form by virtue of the partial gauge fixing \cite{Lopatin:1987hz,Skvortsov:2006at}. For the massless
fields it just reproduces the Fronsdal action \cite{fronsdal}, and for the partially massless fields it
yields  the Lagrangian formulation of Deser and Waldron \cite{pmf}, and Zinoviev \cite{zin}.

\abz The formulation presented below aims to develop a Lagrangian framework that
operates with infinite series of HS symmetric fields in $\ads$ thus providing a starting point for
the study of HS interactions. It originates from a method  applied for constructing
cubic interactions in $AdS_5$ \cite{Vasiliev:2001wa, Alkalaev:2002rq} and involves a description
of HS fields with the help of two sets of auxiliary variables \footnote{The auxiliary variables can
be either vectors or spinors, depending on a particular
realization of HS fields. In our approach we use $o(d-1,2)$ vector variables.}, say, $X$ and $Y$. The action
functional is built then in the following schematic form
\be
\label{AB}
\cS_2 = \int_{\cM^d} \tilde{H}\Big(\frac{\d}{\d Y}\,, \frac{\d}{\d X} \Big) \; R(Y)\;R(X)\Big|_{X=Y=0}\;,
\ee
where $\tilde{H}$ is some differential operator acting on a tensor product of two HS fields.
This scheme has been taken as a pattern for a description of
free mixed-symmetry HS fields in $\ads$ \cite{ASV2}. In the sequel we exploit the main idea of
introducing more than one set of auxiliary variables in the form most close to that of
Ref. \cite{ASV2}.

\abz It is worth remarking that the frame-like action (\ref{AB}) built with the help of the
operator acting on a tensor product of two fields resembles  the typical  way of
introducing interactions in String Field Theory
\footnote{I am grateful to J. Buchbinder and A. Sagnotti for pointing out this to me.} (for review, see \textit{e.g.}, \cite{Zwiebach:1992ie}).
In particular, the SFT inspired approach was used in the higher spin theory for
the study of cubic interactions of massless fields
\cite{Buchbinder:2006eq,Fotopoulos:2007nm,Fotopoulos:2007yq,Fotopoulos:2006ci}.

\vspace{10mm}
\noindent \textit{4.1. $\;\;$ Triple system of auxiliary variables}
\label{sec:actfock}
\vspace{5mm}

\noindent Let us supplement undotted variables introduced in section 3.1 by dotted ones and define a set
$Y_i^A=(Y_\alpha^A, Y_{\dalpha}^B)$, with $\alpha, \dalpha=1,2$, and $A,B=0, ...\, , d$.
Also, we introduce an additional auxiliary anticommuting variable $\theta^A$ that transforms
as $o(d-1,2)$ vector \cite{ASV2}.

\abz The following differential operators
\be
\label{opers1}
\bs^{ij} = \eta^{AB}\frac{\d^2}{\d Y^A_i\d Y^B_j}\;,
\qquad
\bv^i = V^A \frac{\d}{\d Y^A_i}\;
\ee
commute to each other and
\be
\label{opers2}
\ba{c}
\dps
\bta^i = \eta^{AB}\frac{\d^2}{ \d Y^A_i \,\d \theta^B}\;,
\qquad
\chi = V^A\frac{\d}{\d \theta^A}\;,
\qquad
E_0 = E_0^A\frac{\d}{\d \theta^A}
\ea
\ee
anticommute to each other and commute with the set of operators (\ref{opers1}). The combination
\be
\label{opers3}
\Gamma = \frac{1}{(d+1)!}\;\epsilon_{A_1\,\cdots\; A_{d+1}}\,\theta^{A_1}\cdots\theta^{A_{d+1}}\;
\ee
is built with the help of  the $(d+1)$-dimensional Levi-Civita symbol.
It provides a convenient way to work with the Levi-Civita symbol being a part the action (\ref{action0}).

\vspace{10mm}
\noindent \textit{4.2. $\;\;$ Bilinear symmetric form}
\label{sec:actfock}
\vspace{5mm}

\noindent The operators introduced above are the constituents of the following bilinear form
\be
\label{actfock}
\cA(F,G) = \int_{\cM^d} \cH(\bs,\bta,\bv )\, (\wedge E_0)^{d-4}\,\chi\, \Gamma
\wedge F (x|\,Y) \wedge G (x|\,\dot{Y})\Big|_{Y=\dot Y=\theta=0}\;\;\;,
\ee
where  $(\wedge E_0)^k$ stands for $k$-th exterior power of the background frame field, 2-form fields
$F(x|\,Y)$ and $G(x|\,\dot Y)$ are expansions (\ref{denot}) in undotted and dotted variables
$Y_\alpha^A$ and $Y_{\dot \alpha}{}^A$, respectively.  They are subject to the $sp(2)$ invariance
conditions
\be
T_{\alpha\beta}\,F(x|\,Y) = 0\;,
\qquad
T_{\dot \alpha\dot \beta}\,G(x|\,\dot Y) =0 \;.
\ee
The operators (\ref{opers1}), (\ref{opers2}), which are arguments of the function
\be
\label{H}
\cH(\bs,\bta,\bv )
\equiv
\cH(\bs^{\alpha\beta},\bs^{\dot \alpha \dot \beta}, \bs^{\alpha\dot \alpha},
\bv^{\alpha},\bv^{\dot \alpha},\bta^{\alpha},\bta^{\dot \alpha})\;,
\ee
perform contractions of $o(d-1,2)$ indices inside the bilinear form:
$\bs^{\alpha\beta}$ and $\bs^{\dot \alpha \dot \beta}$ take traces of $F(Y)$ and $G(\dot Y)$,
respectively; $\bs^{\alpha\dot \alpha}$ contracts indices from $F(Y)$ and
$G(\dot Y)$; $\bv^{\alpha}$ and $\bv^{\dot \alpha}$ put the compensator $V^A$ on
$F(Y)$ and $G(\dot Y)$; finally, $\bta^{\alpha}$ and $\bta^{\dot \alpha}$ contract indices
of $F(Y)$ and $G(\dot Y)$ with the $o(d-1,2)$ Levi-Civita symbol. Note that the operators (\ref{opers1}) and
(\ref{opers2}) are tensors with respect to the $sp(2)$  transformations.

\abz We require the bilinear form (\ref{actfock}) to be symmetric
\be
\label{sym}
\cA(F,G)=\cA(G,F)\;,
\ee
since for $F=G$ antisymmetric terms disappear anyway.
The symmetry property is equivalent to the invariance under the exchange of dotted and undotted
variables inside the expression (\ref{actfock}). It imposes the following constraints on the
form of the function $\cH$
\be
\label{eulers}
\bs^{\alpha\beta}\frac{\d \cH}{\d \bs^{\alpha\beta}}=\bs^{\dot \alpha\dot \beta}\frac{\d \cH}{\d \bs^{\dot \alpha \dot \beta}}\;,
\qquad
\bv^\alpha\frac{\d \cH}{\d \bv^\alpha}=\bv^{\dot \alpha}\frac{\d \cH}{\d \bv^{\dot \alpha}}\;,
\qquad
\bta^\alpha\frac{\d \cH}{\d \bta^\alpha} = \bta^{\dot \alpha}\frac{\d \cH}{\d \bta^{\dot \alpha}}\;,
\ee
which mean that dotted and undotted operators enter the function $\cH$ in equal portions. Note that the
operators $\bs^{\alpha\dot\beta}$ satisfy  this condition automatically.

\abz Apart from the relations (\ref{eulers}) the function $\cH$ should satisfy
\be
\label{degn}
(\bta_\alpha \frac{\d }{\d \bta_\alpha}+\bta_{\dot \alpha} \frac{\d }{\d \bta_{\dot \alpha}})\cH = 4\, \cH\;.
\ee
The reason is that the bilinear form permits only the maximal number of derivatives in
the anticommuting variables $\theta^A$, \textit{i.e.}, $d+1$,
otherwise it is zero. This fact becomes obvious by virtue of the following relations
\be
\frac{\d}{\d \theta^{A_1}}\cdots \frac{\d}{\d \theta^{A_m}}\, \Gamma\Big|_{\theta=0}
=
\left\{
\ba{l}
\epsilon_{A_1\,\cdots\; A_{d+1}}\,,\;\;\;m=d+1\;,
\\
0\,,\;\qquad\;\;\;\;\;\;\;m\neq d+1\;.
\ea
\right.
\ee
Constituents of the bilinear form that contain derivatives in $\theta^A$ are those listed in (\ref{opers2}).
By definition of the bilinear form, there are $d-4$ derivatives coming from the frame fields $E_0$ and
one coming from the quantity $\chi$. It follows that the remaining four derivatives should come from
the variables $\bta_i$ what justifies the relation (\ref{degn}).

\vspace{7mm}
\noindent \textit{4.3. Auxiliary $sp(2)$ invariant variables}
\label{sec:actfock}
\vspace{5mm}

\noindent Let us now make the following observation. Consider a polynomial
$F(X)$
\be
\dps F(X) = F^{\alpha_1 ... \alpha_n}_{A_1 ... A_n} X_{\alpha_1}^{A_1} \cdots X_{\alpha_n}^{A_n}\;,
\ee
where $X_\alpha^A$ stands for either an auxiliary variable $Y_\alpha^A$ or a
derivative $\dps \frac{\d}{\d Y^\alpha_A}$. The coefficients are obviously   symmetric with
respect to the exchange of pairs $(A, \alpha)$
\be
F^{\alpha_1\, ...\alpha_i ... \alpha_j ... \alpha_n}_{A_1 ... A_i ... A_j ... A_n}
=
F^{\alpha_1 ...\alpha_j ... \alpha_i ... \alpha_n}_{A_1 ... A_j ... A_i ... A_n}\;.
\ee
It follows that if $sp(2)$ indices have a definite Young symmetry type then $o(d-1,2)$
indices have the same symmetry type and vice versa. As an example of using this duality just mention rectangular
$o(d-1,2)$ tableaux that correspond to rectangular $sp(2)$ tableaux, or, by using the
Levi-Civita symbol, to $sp(2)$ singlets, justifying in that way the $sp(2)$ invariance condition (\ref{sl_eq}).
Note also that one of dual symmetries may not be seen manifestly as it happens for $sp(2)$ singlets $F(Y)$
with the expansion coefficients written in non-manifestly covariant $sp(2)$ fashion (\ref{dec_two}).

\abz The coincidence of dual symmetries makes the function $\cH$ to be $sp(2)$ invariant. Indeed,
the function $\cH$ depends on the operators (\ref{opers1}), (\ref{opers2}) that perform contractions of two $o(d-1,2)$ rectangular
tensors. It means precisely that $o(d-1,2)$ indices of these operators should form a group described by
a rectangular tableau. Equivalently, $sp(2)$ indices of the operators should form a rectangular tableau.
In other words, the function $\cH$ should satisfy the $sp(2)$ invariance conditions
\be
\label{Hsp}
T_{\alpha\beta}\cH = T_{\dot \alpha\dot\beta} \cH = 0\;.
\ee
In particular, $sp(2)$ invariance of the fields implies that $\cH$ is defined
modulo contributions proportional to $sp(2)$ generators
\be
\cH \sim \cH+H^{\alpha\beta}T_{\alpha\beta} + H^{\dot \alpha\dot\beta}T_{\dot \alpha\dot\beta}\;,
\ee
where the coefficients in front of the generators are arbitrary symmetric tensors depending
on the same arguments as the function $\cH$.

\abz The expansion coefficients of the function $\cH$  are generically written
as
\be
\label{fun-inv}
\cH_{\alpha_1, \,...\, ,\alpha_{2n};\,\dot \alpha_1, \,...\, ,\dot \alpha_{2n}}
\sim
(\epsilon_{\alpha_1\alpha_2}\cdots \epsilon_{\alpha_{2n-1}\alpha_{2n}})\;
(\epsilon_{\dot\alpha_1\dot\alpha_2}\cdots \epsilon_{\dot\alpha_{2n-1}\dot\alpha_{2n}})\;,
\ee
modulo pre-factors depending on the parameter $n$.
Both dotted and undotted indices enter in equal portions as a corollary
of symmetry conditions (\ref{sym}), (\ref{eulers}) and form a rectangular invariant
representation of $sp(2)$ algebra. The invariance of the coefficients implies
that they are tensor products of the Levi-Civita tensors $\epsilon_{\alpha\beta}$ and $\epsilon_{\dot \alpha\dot \beta}$
\vspace{2mm}

\be
\label{gendecgr2}
\begin{picture}(212,05)(-1,4)
\unitlength=0.6mm
\linethickness{0.210mm}
\put(-5,05){$\Big($}

\put(03,02){\line(1,0){5}} 
\put(03,7){\line(1,0){5}} 
\put(03,12){\line(1,0){5}} 
\put(03,2){\line(0,1){10}} 
\put(08,2){\line(0,1){10}} 

\put(15,05){$\otimes\; \cdots \;\otimes$}

\put(45,02){\line(1,0){5}} 
\put(45,7){\line(1,0){5}} 
\put(45,12){\line(1,0){5}} 
\put(45,2){\line(0,1){10}} 
\put(50,2){\line(0,1){10}} 

\put(17,-2){\footnotesize $n \;\; times$}

\put(54,05){$\dps\Big)_{\rm sym}=$}

\put(95,12){\line(1,0){40}} 
\put(95,2){\line(1,0){40}} 
\put(95,7){\line(1,0){40}} 
\put(110,2){\line(0,1){10}} 
\put(105,2){\line(0,1){10}} 
\put(100,2){\line(0,1){10}} 
\put(95,2){\line(0,1){10}} 
\put(115,2){\line(0,1){10}} 

\put(135,2){\line(0,1){10}} 
\put(130,2){\line(0,1){10}} 
\put(125,2){\line(0,1){10}} 
\put(120,2){\line(0,1){10}} 
\put(115,2){\line(0,1){10}} 

\put(114,-5){\footnotesize $n$}
\put(114,15){\footnotesize $n$}
\put(150,0){$.$}
\end{picture}
\ee

\abz Note that because of the antisymmetrization condition
$\epsilon^{[\alpha\beta} \epsilon^{\gamma]\rho}=0$,  a product of $2d$ Levi-Civita tenors yields  directly a Young tableau
without additional symmetrizations. As a result, the form of the expansion coefficients (\ref{fun-inv}) becomes
obvious.

\abz It is now clear how variables are arranged inside the function $\cH$. They just form various
pairings  with the $2d$ Levi-Civita tenors in an $sp(2)$ invariant fashion. The idea is to single
out those pairings that are elementary in the sense that all other possible pairings are
their combinations. It is easy see that they are given by \footnote{The factor 1/4 in $\textsf{c}_2$ is introduced just for the
convenience in further calculations.}
\be
\textsf{c}_1=\epsilon_{\alpha\beta}\epsilon_{\dot \alpha \dot \beta}\,\bs^{\alpha\dot\alpha} \,\bv^\beta\bv^{\dot \beta}\;,
\qquad
\textsf{c}_2=\frac{1}{4}\,\epsilon_{\alpha\beta}\epsilon_{\dot \alpha \dot \beta}\, \bs^{\alpha\dot\alpha}\,\bs^{\beta\dot\beta}\;,
\ee
and by four more involving the trace annihilation operators $\bs^{\alpha\beta}$ and $\bs^{\dot \alpha\dot \beta}$,
\be
\ba{l}
\textsf{c}_3=(\epsilon_{\alpha\beta}\epsilon_{\gamma\rho}\,\bs^{\alpha\gamma}\bv^\beta\bv^\rho)
(\epsilon_{\dot\alpha\dot\beta}\epsilon_{\dot\gamma\dot\rho}\,\bs^{\dot\alpha\dot\gamma}\bv^{\dot\beta}\bv^{\dot\rho})\;,
\\
\\
\textsf{c}_4 = \epsilon_{\alpha\beta}\epsilon_{\gamma\rho}\,\epsilon_{\dot\alpha\dot\beta}\epsilon_{\dot\gamma\dot\rho}\,
\bs^{\alpha\dot\alpha}\bs^{\gamma\dot\gamma}\bs^{\beta\rho}\bs^{\dot\beta\dot\rho}\;,

\\
\\
\textsf{c}_5 = \epsilon_{\alpha\beta}\epsilon_{\gamma\rho}\,\epsilon_{\dot\alpha\dot\beta}\epsilon_{\dot\gamma\dot\rho}\,
\bs^{\alpha\dot\alpha}\bs^{\beta\rho}\bs^{\dot \beta\dot \rho}\bv^{\gamma}\bv^{\dot \gamma}\;,
\\
\\
\textsf{c}_6 = (\epsilon_{\alpha\beta}\epsilon_{\gamma\rho}\,\bs^{\alpha\gamma}\bs^{\beta\rho})
(\epsilon_{\dot\alpha\dot\beta}\epsilon_{\dot\gamma\dot\rho}\,\bs^{\dot\alpha\dot\gamma}\bs^{\dot\beta\dot\rho})\;.
\ea
\ee
There are also some $sp(2)$ singlet pairings that involve anticommuting variables
$\bta_\alpha$ and $\bta_{\dot \alpha}$. However, the function $\cH$ is required to contain exactly four
anticommuting variables  (\ref{degn}), and the only possible
combination reads

\be
\label{bta}
\bta = (\epsilon_{\alpha\beta}\bta^\alpha\bta^{\beta})(\epsilon_{\dot\alpha\dot\beta}\bta^{\dot\alpha}\bta^{\dot\beta})\;.
\ee
It is obvious that $\bta\bta=0$.

\abz It is convenient to visualize the above contractions by virtue of
the following pictorial representations

\vspace{3mm}

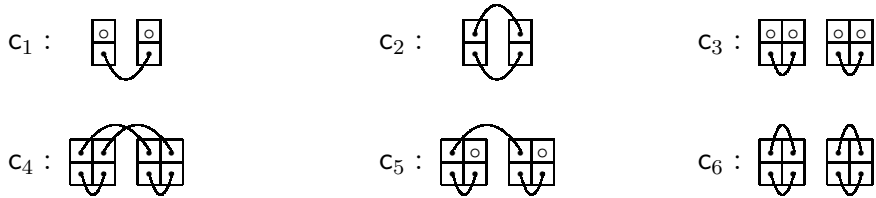
\be
\nonumber
\ba{c}
\textsf{c}_1:
\begin{picture}(50,12)(-12,14.5)
\unitlength=0.6mm

{\linethickness{0.210mm}

\put(00,15){\line(1,0){5}} 
\put(00,10){\line(1,0){5}} 
\put(00,05){\line(1,0){5}} 
\put(00,05){\line(0,1){10}} 
\put(05,05){\line(0,1){10}} 

\put(2.5,12){\circle{2}}
\put(2.5,7.5){\circle*{1}}

\put(10,15){\line(1,0){5}} 
\put(10,10){\line(1,0){5}} 
\put(10,05){\line(1,0){5}} 
\put(10,05){\line(0,1){10}} 
\put(15,05){\line(0,1){10}} 

\put(12.5,12){\circle{2}}
\put(12.5,7.5){\circle*{1}}
}
\qbezier(12.7,7.5)(7,-4)(2.5,7.5)
\end{picture}
\qquad\qquad\qquad
\textsf{c}_2:
\begin{picture}(30,12)(-12,14.5)
\unitlength=0.6mm

{\linethickness{0.210mm}

\put(00,15){\line(1,0){5}} 
\put(00,10){\line(1,0){5}} 
\put(00,05){\line(1,0){5}} 
\put(00,05){\line(0,1){10}} 
\put(05,05){\line(0,1){10}} 

\put(2.5,12){\circle*{1}}
\put(2.5,7.5){\circle*{1}}

\put(10,15){\line(1,0){5}} 
\put(10,10){\line(1,0){5}} 
\put(10,05){\line(1,0){5}} 
\put(10,05){\line(0,1){10}} 
\put(15,05){\line(0,1){10}} 

\put(12.5,12){\circle*{1}}
\put(12.5,7.5){\circle*{1}}
}
\qbezier(12.7,7.5)(7,-4)(2.5,7.5)
\qbezier(12.7,12.5)(7,24)(2.5,12.5)
\end{picture}
\qquad\qquad\qquad
\textsf{c}_3:
\begin{picture}(50,12)(-12,14.5)
\unitlength=0.6mm

{\linethickness{0.210mm}

\put(00,15){\line(1,0){5}} 
\put(00,10){\line(1,0){5}} 
\put(00,05){\line(1,0){5}} 
\put(00,05){\line(0,1){10}} 
\put(05,05){\line(0,1){10}} 

\put(-5,15){\line(1,0){5}} 
\put(-5,10){\line(1,0){5}} 
\put(-5,05){\line(1,0){5}} 
\put(-5,05){\line(0,1){10}} 
\put(0,05){\line(0,1){10}} 

\put(2.5,12){\circle{2}}
\put(2.5,7.5){\circle*{1}}

\put(-2.5,12){\circle{2}}
\put(-2.5,7.5){\circle*{1}}

\put(10,15){\line(1,0){5}} 
\put(10,10){\line(1,0){5}} 
\put(10,05){\line(1,0){5}} 
\put(10,05){\line(0,1){10}} 
\put(15,05){\line(0,1){10}} 

\put(15,15){\line(1,0){5}} 
\put(15,10){\line(1,0){5}} 
\put(15,05){\line(1,0){5}} 
\put(15,05){\line(0,1){10}} 
\put(20,05){\line(0,1){10}} 

}

\put(12.5,12){\circle{2}}
\put(12.5,7.5){\circle*{1}}

\put(17.5,12){\circle{2}}
\put(17.5,7.5){\circle*{1}}

\qbezier(17.7,7.5)(15,-2)(12.5,7.5)

\qbezier(2.7,7.5)(0,-2)(-2.5,7.5)

\end{picture}
\\
\\
\\
\textsf{c}_4:
\begin{picture}(50,12)(-12,14.5)
\unitlength=0.6mm

{\linethickness{0.210mm}

\put(00,15){\line(1,0){5}} 
\put(00,10){\line(1,0){5}} 
\put(00,05){\line(1,0){5}} 
\put(00,05){\line(0,1){10}} 
\put(05,05){\line(0,1){10}} 

\put(-5,15){\line(1,0){5}} 
\put(-5,10){\line(1,0){5}} 
\put(-5,05){\line(1,0){5}} 
\put(-5,05){\line(0,1){10}} 
\put(0,05){\line(0,1){10}} 

\put(2.5,12){\circle*{1}}
\put(2.5,7.5){\circle*{1}}

\put(-2.5,12){\circle*{1}}
\put(-2.5,7.5){\circle*{1}}

\put(10,15){\line(1,0){5}} 
\put(10,10){\line(1,0){5}} 
\put(10,05){\line(1,0){5}} 
\put(10,05){\line(0,1){10}} 
\put(15,05){\line(0,1){10}} 

\put(15,15){\line(1,0){5}} 
\put(15,10){\line(1,0){5}} 
\put(15,05){\line(1,0){5}} 
\put(15,05){\line(0,1){10}} 
\put(20,05){\line(0,1){10}} 

}

\put(12.5,12){\circle*{1}}
\put(12.5,7.5){\circle*{1}}

\put(17.5,12){\circle*{1}}
\put(17.5,7.5){\circle*{1}}

\qbezier(17.7,7.5)(15,-2)(12.5,7.5)
\qbezier(17.5,12.5)(10,24)(2.5,12.5)

\qbezier(2.7,7.5)(0,-2)(-2.5,7.5)
\qbezier(12.7,12.5)(5,24)(-2.5,12.5)

\end{picture}
\qquad\qquad\qquad
\textsf{c}_5:
\begin{picture}(30,12)(-12,14.5)
\unitlength=0.6mm

{\linethickness{0.210mm}

\put(00,15){\line(1,0){5}} 
\put(00,10){\line(1,0){5}} 
\put(00,05){\line(1,0){5}} 
\put(00,05){\line(0,1){10}} 
\put(05,05){\line(0,1){10}} 

\put(-5,15){\line(1,0){5}} 
\put(-5,10){\line(1,0){5}} 
\put(-5,05){\line(1,0){5}} 
\put(-5,05){\line(0,1){10}} 
\put(0,05){\line(0,1){10}} 

\put(2.5,12){\circle{2}}
\put(2.5,7.5){\circle*{1}}

\put(-2.5,12){\circle*{1}}
\put(-2.5,7.5){\circle*{1}}

\put(10,15){\line(1,0){5}} 
\put(10,10){\line(1,0){5}} 
\put(10,05){\line(1,0){5}} 
\put(10,05){\line(0,1){10}} 
\put(15,05){\line(0,1){10}} 

\put(15,15){\line(1,0){5}} 
\put(15,10){\line(1,0){5}} 
\put(15,05){\line(1,0){5}} 
\put(15,05){\line(0,1){10}} 
\put(20,05){\line(0,1){10}} 

}

\put(12.5,12){\circle*{1}}
\put(12.5,7.5){\circle*{1}}

\put(17.5,12){\circle{2}}
\put(17.5,7.5){\circle*{1}}

\qbezier(17.7,7.5)(15,-2)(12.5,7.5)

\qbezier(2.7,7.5)(0,-2)(-2.5,7.5)
\qbezier(12.7,12.5)(5,24)(-2.5,12.5)

\end{picture}
\qquad\qquad\qquad
\textsf{c}_6:
\begin{picture}(50,12)(-12,14.5)
\unitlength=0.6mm

{\linethickness{0.210mm}

\put(00,15){\line(1,0){5}} 
\put(00,10){\line(1,0){5}} 
\put(00,05){\line(1,0){5}} 
\put(00,05){\line(0,1){10}} 
\put(05,05){\line(0,1){10}} 

\put(-5,15){\line(1,0){5}} 
\put(-5,10){\line(1,0){5}} 
\put(-5,05){\line(1,0){5}} 
\put(-5,05){\line(0,1){10}} 
\put(0,05){\line(0,1){10}} 

\put(2.5,12){\circle*{1}}
\put(2.5,7.5){\circle*{1}}

\put(-2.5,12){\circle*{1}}
\put(-2.5,7.5){\circle*{1}}

\put(10,15){\line(1,0){5}} 
\put(10,10){\line(1,0){5}} 
\put(10,05){\line(1,0){5}} 
\put(10,05){\line(0,1){10}} 
\put(15,05){\line(0,1){10}} 

\put(15,15){\line(1,0){5}} 
\put(15,10){\line(1,0){5}} 
\put(15,05){\line(1,0){5}} 
\put(15,05){\line(0,1){10}} 
\put(20,05){\line(0,1){10}} 

}

\put(12.5,12){\circle*{1}}
\put(12.5,7.5){\circle*{1}}

\put(17.5,12){\circle*{1}}
\put(17.5,7.5){\circle*{1}}

\qbezier(17.7,7.5)(15,-2)(12.5,7.5)
\qbezier(17.5,12.5)(15,24)(12.5,12.5)

\qbezier(2.7,7.5)(0,-2)(-2.5,7.5)
\qbezier(2.7,12.5)(0,24)(-2.5,12.5)

\end{picture}

\ea
\ee

\vspace{5mm}

\abz and

\be
\nonumber
\bta:
\begin{picture}(50,12)(-25,14.5)
\unitlength=0.6mm

{\linethickness{0.210mm}

\put(00,15){\line(1,0){5}} 
\put(00,10){\line(1,0){5}} 
\put(00,05){\line(1,0){5}} 
\put(00,05){\line(0,1){10}} 
\put(05,05){\line(0,1){10}} 

\put(2.5,12){\circle*{1}}
\put(2.5,7.5){\circle*{1}}

\put(10,15){\line(1,0){5}} 
\put(10,10){\line(1,0){5}} 
\put(10,05){\line(1,0){5}} 
\put(10,05){\line(0,1){10}} 
\put(15,05){\line(0,1){10}} 

\put(12.3,12){\circle*{1}}
\put(12.3,7.5){\circle*{1}}

\put(12.5,10.5){- - -}
\put(12.5,6){- - -}

\put(-8.7,10.5){- - -}
\put(-8.7,6){- - -}
}
\end{picture}
\ee

\abz Here two-cell columns are the Levi-Civita tensors, the arcs denote cross-contractions
$\bs^{\alpha\dot\alpha}$ and traces $\bs^{\alpha\beta}, \bs^{\dot\alpha\dot \beta}$,
the circles denote $\bv^\alpha$ and $\bv^{\dot\alpha}$, and the dashed  lines denote anticommuting $\bta^\alpha$ and $\bta^{\dot\alpha}$.

\abz The coinciding dual Young symmetry types make the above graphs applicable directly for $o(d-1,2)$ tableaux.
It is remarkable that two Young symmetry bases  are involved simultaneously, a symmetric one,
for $o(d-1,2)$ tableaux, and an antisymmetric one, for $sp(2)$ tableaux.
It follows that some symmetry properties not seen directly in one basis become clear in another one.

\abz The general form of the function $\cH$ written as a power series in the
new $sp(2)$ invariant variables is
\be
\label{newfunH}
\cH=H(\textsf{c})\, \bta\;,
\qquad
H(\textsf{c})=\sum_{k_i\geq 0}^{\infty} \xi(k_i\,;d) \prod_{i=1}^6 (\textsf{c}_i)^{k_i}\,\;,
\ee
where $\xi(k_i\,;d)$ are some $d$-dependent coefficients.

\vspace{10mm}
\noindent {\large\textsc{5. Action for symmetric HS fields}}
\label{sec:act}
\vspace{5mm}

\abz So far we have elaborated the framework that utilizes $sp(2)$ doublets of auxiliary vector
variables for a description of non-symmetric higher-rank tensors. Now we are in a position to
adjust it for Lagrangian formulation of HS field dynamics along the lines discussed in the
beginning of the previous section.

\abz Prior to continue let us make a comment that the $sp(2)$ invariance
condition does not specify particular lengths of Young tableau and just
requires it to be a rectangular block. In particular, it allows one to consider
infinite sets of the same symmetry type tensors  on equal footing. On the contrary,
in the manifest antisymmetric basis for rectangular Young tableaux Howe dual algebra becomes
$sl(m)$, where $m$ is a length
of the uppermost row \cite{howe,Vasiliev:2004cm}. However, the use of antisymmetric basis
is inconvenient because one should introduce then an infinite chain of Howe dual algebras that correspond to
an infinite set of Young tableaux with increasing lengths.

\vspace{10mm}
\noindent \textit{5.1. $\;\;$ Action functional: general properties }
\label{sec:actfock}
\vspace{5mm}

\abz Let us introduce the $sp(2)$ invariant 1-form gauge field
\be
\label{gaugefield}
\Omega(x|\,Y) = \textrm{d}x^{\underline{n}}\, \Omega_{\underline{n}}(x|\,Y)\;,
\qquad
T_{\alpha\beta}\Omega(x|\,Y)=0\;.
\ee
Frame-like higher spin  gauge fields are then identified  with the expansion coefficients of $\Omega(x|\,Y)$ with respect to
the auxiliary variables. An irreducible field (massless or
partially massless) of a given spin $s^\prime=s$ and depth $t^\prime=s-2t$ appears in  $\Omega(x|\,Y)$ in infinitely
many copies because the trace decomposition (\ref{infexp}) requires
\be
\label{Gauge_infexp}
\Omega(Y|\,x) = \sum_{n,\,s,\,t=0}^{\infty}\;
\rho(s,t,n)\, Z_+^n\, \Omega_{s,\, t;\,n}(Y|\,x)\;,
\ee
where $\rho(s,t,n)$ are some normalization coefficients. Note that the parameter
$t$ is necessarily even, and therefore the partially massless fields appearing in (\ref{Gauge_infexp}) are
not arbitrary. On the contrary, massless fields ($t=0$) appear with any value of  spin $s$ from zero to
infinity.

\abz The  field (\ref{gaugefield}) is a gauge connection of the HS algebra
$\hc$, frequently regarded as the off-shell algebra
\cite{Vasiliev:2003ev,Vasiliev:2004cm,Sagnotti:2005ns, Vasiliev:2005zu}. The presence of partially
massless states in the spectrum violates  the unitarity already on the free field level. To get rid
of them a procedure is required that allows one to do this consistently in the sense that the remaining
fields should be organized in a  multiplet of some HS algebra. In $d$ dimensions it
precisely corresponds to the factoring out an ideal of $\hc$ generated by traces of the field
(\ref{gaugefield}). It gives rise to the on-shell algebra $\hs$ of \cite{Vasiliev:2003ev}.
The resulting theory will be described by (\ref{Gauge_infexp}) with all $\Omega_{s,\, t;\,n}(Y|\,x)$
at $t\neq 0, n\neq 0$ set to zero. In $d=5$ a weaker truncation is possible
which drops out all partially massless fields but retains massless fields in infinitely many copies.
The final set of fields corresponds to the spectrum of $5d$ algebra $hu(1,1|8)$
\cite{FradkinLin,Sezgin:2001zs,Vasiliev:2001zy,Vasiliev:2001wa}.

\abz Now the HS  action functional can be defined by virtue of
the bilinear form (\ref{actfock}) in the following way
\be
\label{fin_act}
\cS_2[\Omega] = \half \,\cA(R,R)\;,
\ee
where the linearized curvatures (\ref{R1A}), (\ref{pcurv})
\be
R(Y|\,x) = D_0 \Omega(Y|\,x)
\qquad
\textrm{ and}
\qquad
R(\dot Y|\,x) = D_0 \Omega(\dot Y|\,x)\;
\ee
are associated with  gauge fields (\ref{Gauge_infexp}) and invariant under
the transformations (\ref{litr}), (\ref{pgauge})
\be
\delta \Omega(Y|\,x) = D_0\varepsilon(Y|\,x)
\qquad
\textrm{ and}
\qquad
\delta \Omega(\dot Y|\,x) = D_0\varepsilon(\dot Y|\,x)\;.
\ee
Both curvatures and  gauge parameters naturally inherit the property of $sp(2)$ invariance
\be
\ba
{c}
T_{\alpha\beta}\,R(Y|\,x) = T_{\dot\alpha\dot\beta}\, R(\dot Y|\,x) = 0\;,
\\
\\
T_{\alpha\beta}\,\varepsilon(Y|\,x) = T_{\dot\alpha\dot\beta}\, \varepsilon(\dot Y|\,x) = 0\;.
\ea
\ee

\abz As discussed above, the gauge field $\Omega(Y|\,x)$ describes an infinite sum of
irreducible fields in infinitely many copies. However, it should not be taken for granted that the action functional
(\ref{fin_act}) describes a direct sum of the actions for irreducible fields. In other words, an
additional condition should be imposed that makes the action (\ref{fin_act}) diagonal.
Such a condition requires all the cross-terms containing products of fields
$\Omega_{s,\, t;\,m}(x)$ and $\Omega_{s, \,t;\,n}(x)$ for $m\neq n$ to vanish
\be
\label{diag}
\cS_2[\Omega] = \sum_{n}^{} \sum_{s,\,t}\,\chi(s,\, t;\,n)\,\cS_2[\Omega_{s,\, t;\,n}]\;,
\ee
where $\chi(s,\, t;\,n)$ are some normalization coefficients.
The diagonalization  condition will be considered elsewhere \cite{Alk_next}.

\abz Another condition to be imposed on the action $\cS_2[\Omega]$ is the decoupling
of extra fields \cite{V1,Lopatin:1987hz, Vasiliev:2001wa,Skvortsov:2006at}. It requires that the
extra fields should enter the action through total derivatives, \textit{i.e.},
\be
\label{dec_con}
\frac{\delta \cS_2[\Omega]}{\delta \Omega^{extra}} \equiv 0\;,
\ee
and means that the action depends non-trivially on the physical and the auxiliary fields only.

\abz The action functional (\ref{fin_act}) subject to the diagonalization condition (\ref{diag}) and the extra field
decoupling condition (\ref{dec_con}) describes both massless and partially massless free symmetric
fields and each field appears in infinitely many copies. These two conditions fix the action unambiguously up to
an arbitrary normalization coefficient in front of a given spin action $\cS_2[\Omega_{s,\, t;\,n}]$ (\ref{diag}).

\vspace{10mm}
\noindent \textit{5.2. $\;\;$ Action functional: a non-degenerate set of fields}
\label{sec:actfock}
\vspace{5mm}

\abz In this section we study the general form of the action functional which describes
a non-degenerate set of symmetric fields, that is each field enters in a single copy. In this case the diagonalization condition (\ref{diag}) is relaxed.

\abz Let us suppose that all irreducible fields enter $\Omega(Y|\,x)$ in a single copy and for
this occasion introduce a notation $\Omega_{s, \,t}(Y|\,x)\equiv\Omega_{s, \,t;\,0}(Y|\,x)$.
As discussed in section 3.2, the field  $\Omega_{s, \,t}(Y|\,x)$ satisfies the trace condition
\be
(\bs^{\alpha\beta})^{t+1}\,\Omega_{s,\, t}(Y|\,x) = 0\;.
\ee
Because traces enter $\Omega_{s,\, t}(Y|\,x)$ in a totally symmetric combination (\ref{newF}),
it follows that a non-symmetric combination of the annihilation trace operators acts on it by zero
\be
\label{nonsymtr}
(\epsilon_{\alpha\beta}\epsilon_{\gamma\rho}\,\bs^{\alpha\gamma}\bs^{\beta\rho})\,
\Omega_{s, \,t}(Y|\,x) = 0\;.
\ee
Indeed, the symmetry types of components arising in a tensor product of the trace annihilation
operators $\bs^{\alpha\beta}$ are precisely described by the decomposition (\ref{gendecgr}) that
up to pre-factors takes now
the following form
\be
(\bs^{\alpha\beta})^{n} \sim \sum_{2l+k=n}\,\bs^{\alpha_1 ... \alpha_{2k}}\, Z_-^l\;,
\ee
where the notation are introduced
\be
\bs^{\alpha_1 ... \alpha_{2k}} = \bs^{(\alpha_1\alpha_2} \cdots \bs^{\alpha_{2k-1}\alpha_{2k})}
\qquad
{\rm and}
\qquad
Z_-=\epsilon_{\alpha\beta}\epsilon_{\gamma\rho}\,\bs^{\alpha\gamma}\bs^{\beta\rho}\;.
\ee
Then the result is that the function $H(\textsf{c})$ (\ref{newfunH}) becomes independent of the
variable $\textsf{c}_6$, \textit{i.e.},
\be
\frac{\d H}{\d \textsf{c}_6} = 0\;,
\ee
since $\textsf{c}_6=Z_-{\dot Z}_-$ acts trivially on the fields $\Omega_{s, \,t}(Y|\,x)$ and $\Omega_{s, \,t}(\dot Y|\,x)$.
Another  consequence of the relation (\ref{nonsymtr}) is that modulo terms containing non-symmetric
combinations of traces, the following algebraic constraint takes place
\be
\textsf{c}_5^2 = \textsf{c}_3\textsf{c}_4\;.
\ee
As a result, the function $H$ becomes linear in variable $\textsf{c}_5$, that is
\be
H= H_1(\textsf{c}) + \textsf{c}_5\,H_2(\textsf{c})\;,
\qquad
\frac{\d H_{1}(\textsf{c})}{\d \textsf{c}_5}=\frac{\d H_{2}(\textsf{c})}{\d \textsf{c}_5}=0\;.
\ee
We see that the function $H(\textsf{c})$ corresponding to the infinite series  of irreducible symmetric
fields (each in a single copy) depends essentially on four variables. This fact is in agreement with the component form of HS actions
originally elaborated in \cite{Lopatin:1987hz,Vasiliev:2001wa, Skvortsov:2006at}. Indeed, before
the extra field decoupling condition is imposed, the coefficients in the action are
parameterized by four numbers, $a(s,t; m,l)$, where $s$ and $t$ define a size of traceless $o(d-1,2)$
Young tableau, $m$ and $l$ correspond to the numbers of the compensator vectors contracted with first
and second rows, respectively. The massless fields are described by rectangular tableaux and therefore
one is left with just two parameters, $s$ and $m$.
Within our approach it corresponds to
\be
H=H(\textsf{c}_1, \textsf{c}_2)\;,
\ee
where  all traceful  contributions are eliminated because  the variables involving trace operators act
trivially, $\d H/\d \textsf{c}_{i}=0$, $i=3,4,5,6$.

\vspace{10mm}
\noindent \textit{5.3. $\;\;$ Action for massless symmetric fields}
\label{sec:actfock}
\vspace{5mm}

\noindent We have emphasized earlier that in order to have unitary dynamics
partially massless fields are required to decouple from the whole system of fields.
In this section we drop out all partially massless fields by hand and consider
massless fields only. Such a truncation
is consistent on the free field level. The resulting set of fields with spins $0\leq s < \infty$ form
the multiplet of symmetric massless fields of the algebra $\hs$ \cite{Vasiliev:2004cm}. Note that lower spin fields
with $s\leq 1$ do not admit a frame-like Lagrangian form (\ref{action0}) and should be described by standard Klein-Gordon and Maxwell actions.

\abz The action for a single massless field of spin $s$ is written down as
\be
\label{var}
\cS_2[\Omega_{s,\,0}] = \half\int_{\cM^d}  \cH \, (\wedge E_0)^{d-4}\,\chi\, \Gamma
\wedge R_{s,\,0} (x|\,Y) \wedge R_{s,\,0} (x|\,\dot Y)\Big|_{Y=\dot Y=0}\;,
\ee
where
\be
\label{funct_massless}
\cH=H(\textsf{c}_1,\textsf{c}_2)\, \bta\;,
\qquad
H(\textsf{c}_1,\textsf{c}_2) = \frac{1}{(s-1)}\;\sum_{m=0}^{s-2}\xi(m;d,s)\; \textsf{c}_1^{s-m-2}\,\textsf{c}_2^m\;.
\ee
Here $\xi(m;d,s)$ are arbitrary real coefficients parameterized by three numbers, fixed $s\geq 2$ and $d\geq 4$, and
running $m$.

\abz To impose the extra field decoupling condition we make use of  the method elaborated
in Ref. \cite{ASV2}. Namely, one observes that in order to have a manifest gauge invariance,
the action is always  written down with the extra fields, at least formally \footnote{Having decoupled
extra fields, the action can be cast into a minimal form with just two fields, the physical and the auxiliary,
but the residual gauge invariance is implicit \cite{Vasiliev:1980as,Boulanger:2003vs}.}. On the contrary,
the manifestly gauge invariant field equations satisfying the decoupling condition are easier  to find since they depend
on two fields only, physical and auxiliary ones. The idea is then to reconstruct the action from
the known field equations by requiring them to follow from the action. To perform a reconstruction the cohomological technique based
on the so-called $\cQ$-complex was elaborated \cite{ASV2}. Here we find
the action from the field equations by another method, which turns out to be more appropriate for our purposes.

\abz The field equations that follow from a variation of the action for massless fields
and satisfy the extra field decoupling condition
have the form
\be
\label{var}
\delta\cS_2[\Omega_{0,\,s}] = \int_{\cM^d}  \cE(\bs,\bta,\bv) \, (\wedge E_0)^{d-3}\,\chi\, \Gamma
\wedge R_{s,\,0} (x|\,Y) \wedge \delta \Omega_{s,\,0} (x|\,\dot Y)\Big|_{Y=\dot Y=0}=0\;
\ee
and the function $\cE(\bs,\bta,\bv)$ is defined in the $sp(2)$ invariant way as
\be
\label{equat}
\cE= (\bar\pi  \dot {\bar\tau}-\dot {\bar\pi}   {\bar\tau})\, T(\textsf{c}_1)\;,
\ee
where  variables $\bar\pi$ and $\bar\tau$ and their dotted cousins are the following $sp(2)$ singlet pairings
\be
\ba{l}
\bar\pi = \epsilon_{\alpha\beta}\bta^\alpha\bta^\beta\;,
\quad
\bar\tau = \epsilon_{\alpha\beta}\bv^\alpha\bta^\beta\;,
\\
\\
\dot{\bar\pi} = \epsilon_{\dot\alpha\dot\beta}\bta^{\dot\alpha}\bta^{\dot\beta}\;,
\quad
\dot{\bar\tau} = \epsilon_{\dot\alpha\dot\beta}\bv^{\dot\alpha}\bta^{\dot\beta}\;.
\ea
\ee
Note that $\bar\tau\bar\tau=0$ and
$\dot{\bar\tau}\dot{\bar\tau}=0$, and ${\bar\pi}\dot{\bar\pi}=\bta$ (cf. (\ref{bta})). The function $T(\textsf{c}_1)$ is an arbitrary polynomial,
which in the case of a single massless field becomes a monomial
\be
\label{funcT}
T(\textsf{c}_1) = \textsf{c}_1^{s-2}\;.
\ee
The extra field decoupling condition is automatically  satisfied by the field equations (\ref{var}).
Indeed, according to formula (\ref{phaux}) the physical and the auxiliary fields are contained
in the original field $\Omega_{s,\,0}(x|\,\dot Y)$ contracted with $(s-2)$ compensators.
It explains the appearance of the function $T(\textsf{c}_1)$ (\ref{funcT}).
Then, the first term of (\ref{equat}) contains maximal possible number $(s-1)$ of compensators
contracted with $\delta\Omega_{s,\,0}(x|\,\dot Y)$ and therefore, as discussed in section 2, corresponds to
the variation with respect to the physical field.
Analogously, the second term contains $(s-2)$ compensators contracted with
$\delta\Omega_{s,\,0}(x|\,\dot Y)$ and therefore,
corresponds to the variation with respect to the auxiliary field.
In both the contractions, the remaining index of
$R_{s,\,0}(x|\,Y)$ or $\delta\Omega_{s,\,0}(x|\,\dot Y)$
is contracted with the $o(d-1,2)$ Levi-Civita tensor. The terms $(\bar\pi  \dot {\bar\tau}-\dot {\bar\pi}   {\bar\tau})$ in (\ref{equat})
can be described by the following pictorial representation
\be
\nonumber
\begin{picture}(50,12)(-25,14.5)
\unitlength=0.6mm

{\linethickness{0.210mm}

\put(00,15){\line(1,0){5}} 
\put(00,10){\line(1,0){5}} 
\put(00,05){\line(1,0){5}} 
\put(00,05){\line(0,1){10}} 
\put(05,05){\line(0,1){10}} 

\put(2.5,12){\circle*{1}}
\put(2.5,7.5){\circle*{1}}

\put(10,15){\line(1,0){5}} 
\put(10,10){\line(1,0){5}} 
\put(10,05){\line(1,0){5}} 
\put(10,05){\line(0,1){10}} 
\put(15,05){\line(0,1){10}} 

\put(12.3,12){\circle*{1}}

\put(12.5,10.5){- - -}
\put(12.5,7){\circle{2}}

\put(-8.7,10.5){- - -}
\put(-8.7,6){- - -}

}

\end{picture}
\qquad
\begin{picture}(50,12)(-25,14.5)
\unitlength=0.6mm

{\linethickness{0.210mm}

\put(00,15){\line(1,0){5}} 
\put(00,10){\line(1,0){5}} 
\put(00,05){\line(1,0){5}} 
\put(00,05){\line(0,1){10}} 
\put(05,05){\line(0,1){10}} 

\put(2.5,12){\circle*{1}}
\put(2.5,7.5){\circle{2}}

\put(10,15){\line(1,0){5}} 
\put(10,10){\line(1,0){5}} 
\put(10,05){\line(1,0){5}} 
\put(10,05){\line(0,1){10}} 
\put(15,05){\line(0,1){10}} 

\put(12.3,12){\circle*{1}}
\put(12.3,7.5){\circle*{1}}

\put(12.5,10.5){- - -}
\put(12.5,6){- - -}

\put(-8.7,10.5){- - -}
}
\end{picture}
\ee
where left and right columns correspond to the curvature and the field variation.

\abz Let us now obtain the variation of the action and equate it to the field equations defined
by (\ref{var}) and (\ref{equat}). It results in the equation
\be
\label{dec_eq}
(2(d-3)+4\textsf{c}_1\frac{\d}{\d \textsf{c}_1}-4\textsf{c}_2\frac{\d}{\d \textsf{c}_1})H(\textsf{c}_1,\textsf{c}_2) = T(\textsf{c}_1)\;.
\ee
It is solved by the following integral expression (see Appendix A.1 for more details)
\be
\label{dec_sol}
H(\textsf{c}_1,\textsf{c}_2)
= \frac{1}{4}\int_0^1 {\rm d} t\; t^{(d-5)/2}\, \exp{\bigg(\frac{1-t}{t}\;\textsf{c}_2\frac{\d}{\d \textsf{c}_1}\bigg)}\;T(t \textsf{c}_1)\;.
\ee
This formula trivially generalizes to the case of an arbitrary polynomial function $T(\textsf{c}_1)$ thus giving
rise to a direct sum of spin-$s$ actions with $2\leq s\leq \infty$.

\abz After some simple algebra (see Appendix A.2) one obtains the expansion coefficients (\ref{funct_massless})
expressed in terms of the beta functions
\be
\label{answer}
\xi(m; d,s)=  \frac{B(m+1,s-m-1+(d-5)/2)}{B(m+1, s-m-1)}\;\;.
\ee
This answer can be compared with the coefficients arising in the component form
of the action. To this end,
one violates manifest $sp(2)$ invariance and introduces instead of $\textsf{c}_1$ and $\textsf{c}_2$
new variables $x_1 = \bs^{1\dot1}\bv^2\bv^{\dot2}$ and $x_2 = \bs^{1\dot1}\bs^{2\dot2}$
that perform row-to-row contractions of two rectangular tableaux. A change of variables done inside
the action gives the expression
\be
c_1^m c_2^n \sim 2^n (m+1)(m+n+1) x_1^m x_2^n\;,
\ee
where $\sim$ means that the equality is valid up to terms proportional to Young symmetrizers
(\ref{Y1}) that trivialize when acting on Young tableaux.
Then, by using formulas of Appendix A.2,
the function (\ref{answer}) can be cast into the more traditional form with (double) factorials
known from Ref. \cite{Vasiliev:2001wa}
\be
\zeta(m; d,s) = \zeta(d,s) \,\frac{(s-m-1)(d-5+2(s-m-2))!!}{(s-m-2)!}\;,
\ee
where $\zeta(d,s)$ is an overall factor in front of the spin-$s$ action.

\abz In conclusion, let us make a comment that for a single massless fields the
Howe dual $sp(2)$ algebra is enhanced to $sp(4)$ but it seems that within our approach $sp(4)$ does not play
any essential role. Therefore it would be interesting to see more deeper implementation
of $sp(4)$ symmetry for the frame-like formulation of symmetric fields. For example, within the first-quantized BRST approach to higher spin dynamics
\cite{Barnich:2004cr} Howe dual $sp(4)$ algebra appears as an extension of first-class
constraint algebra that describes  classical mechanics of a particle with the spin degrees of
freedom.

\vspace{10mm}
\noindent {\large\textsc{6. Conclusions}}
\label{sec:cub}
\vspace{5mm}

\noindent We offered  the new perspective on using $sp(2)$ symmetry in the Lagrangian HS dynamics of
bosonic symmetric fields and elaborated on the idea of introducing $sp(2)$ invariant variables.
The whole formulation is designed to deal with infinite multiplets
of fields naturally appearing as the HS gauge connections.
The present paper can be considered as a first step towards
the study of Lagrangian form of the  HS interactions. Having this in mind let us now summarize our results.
\begin{itemize}

\item We have elaborated more on HS fields written in terms of $sp(2)$ auxiliary variables and,
in particular, studied the trace decompositions that allows one to control contributions of massless and partially massless symmetric fields.
Higher rank tensors described as polynomials  of auxiliary commuting variables $Y_{\alpha}^A$
are in fact elements of the $*$-product algebra generated by
$Y_{\alpha}^A*Y_{\beta}^B-Y_{\beta}^B*Y_{\alpha}^A=\epsilon_{\alpha\beta}\eta^{AB}$.  In this form
these higher rank tensors are naturally appear as the HS connections \cite{Vasiliev:2003ev}.

\item We have introduced the bilinear form $\cA(F,G)$ defined on arbitrary $sp(2)$ singlet fields
$F(Y\,|x)$ and $G(\dot Y\,|x)$.  It serves as the main building block of the HS
action functionals to be considered both on the free field and the interaction levels. The main
novel ingredient is the use of $sp(2)$ singlet variables that allows one to avoid dealing with
the explicit Young symmetrizers inside the bilinear form and thus considerably simplify calculations.

\item Within our approach we have analyzed the general properties of the quadratic higher spin actions and,
in particular, considered the action describing a single symmetric field.  We have explicitly built  the
action for free massless symmetric fields and  reproduced  the well-known expression for
coefficients of the component form of HS action originally obtained in \cite{Vasiliev:2001wa}. Our
answer is given as an integral of the exponential operator and we expect that this form of
coefficients is suitable for dealing with $*$-product when studying the HS interactions.

\item Summarizing the above, our main result is that we have brought together formulations used
previously for the HS algebra \cite{Vasiliev:2003ev} and the HS action functionals
\cite{Vasiliev:2001wa,ASV2} and provided for them a unified framework.

\end{itemize}

\abz Within our approach we could also reproduce the component form of HS action for partially
massless fields \cite{Skvortsov:2006at}. Let us stress once again that partially massless
fields appearing in the trace decompositions of traceful rectangular fields have an
even difference of lengths therefore their spins $s$ and depths $t$ are not arbitrary.

\abz To conclude, let us mention the following directions for the further
study:

\begin{itemize}

\item A Lagrangian form of non-linear dynamics of partially massless and massless fields interacting between
themselves and with the gravity. It is expected that such a theory should be governed by
the "off-shell" algebra $\hc$ \cite{Vasiliev:2004cm}.

\item A Lagrangian form of non-linear dynamics of massless fields only. This theory
is based on the "on-shell" algebra $\hs$, which is the quotient of the "off-shell" algebra \cite{Vasiliev:2003ev,Vasiliev:2004cm}.
On the level of the equations of motion it was formulated in Ref. \cite{Vasiliev:2003ev}.
To develop a Lagrangian formulation we suggest to use the projection technique as we
described in the Introduction.

\end{itemize}

\abz Hopefully, the interaction problems listed above could be explicitly
analyzed at least in the cubic approximation \cite{Alk_next}.

\vspace{10mm}
\noindent \textbf{Acknowledgements}
\vspace{2mm}

\noindent Even this humble work would not ever be possible without my wife Sasha Tertel
and I am thankful to her for all her love and moral support.

\vspace{3mm}

\noindent I am grateful to Augusto Sagnotti for his warm hospitality at La Scuola
Normale Superiore di Pisa, where this work was completed. Many useful
discussions with Maxim Grigoriev, Slava Didenko, Ruslan Metsaev, Oleg Shaynkman,
Zhenya Skvortsov, and Mikhail
Vasiliev are appreciated.

\vspace{3mm}

\noindent The work is partially supported by RFBR grant Nr 05-02-17654, LSS grant Nr 4401.2006.2,
Post Doctoral Fellowship grant INTAS Nr 05-109-4761, and the Alexander von Humboldt Foundation Grant PHYS0167.


\vspace{10mm}
\noindent {\large\textsc{Appendix}} 
\setcounter{equation}{0}\renewcommand{\theequation}{A.\arabic{equation}}
\label{appendix}
\vspace{5mm}

\noindent {\textsf{A.1.}} The extra field decoupling condition
results in the partial differential equation (\ref{dec_eq}) which is a particular case of

$$
\Big(\alpha+\beta \textsf{c}_1\frac{\d}{\d \textsf{c}_1}+\gamma \textsf{c}_2\frac{\d}{\d \textsf{c}_1}\Big)H(\textsf{c}_1,\textsf{c}_2) = T(\textsf{c}_1,\textsf{c}_2)\;,
$$
where $\alpha, \beta$ and $\gamma$ are some constants and $T(\textsf{c}_1,\textsf{c}_2)$ is a given polynomial function. To find
the function $H(\textsf{c}_1, \textsf{c}_2)$ we solve this equation in two steps. First, one observes that the following identity is valid
$$
\exp\Big(\rho\, \textsf{c}_2\frac{\d}{\d c_1}\Big) N_1-N_1 \exp\Big(\rho\, \textsf{c}_2\frac{\d}{\d \textsf{c}_1}\Big) =
\rho \textsf{c}_2\frac{\d}{\d c_1} \exp\Big(\rho\, \textsf{c}_2\frac{\d}{\d \textsf{c}_1}\Big)\;,
$$
where  $\rho$ is an arbitrary constant and $\dps N_1=\textsf{c}_1\frac{\d}{\d \textsf{c}_1}$ is the Euler operator.
By making use of this identity the original equation is cast into the form
$$
\Big(\frac{\alpha}{\beta}+N_1\Big)\tilde H(\textsf{c}_1,\textsf{c}_2)=\frac{1}{\beta}\;\tilde T(\textsf{c}_1,\textsf{c}_2)\;,
$$
where the tildes mark functions transformed as
$$\dps \tilde F(\textsf{c}_1,\textsf{c}_2)=\exp(-\frac{\gamma}{\beta}\,\textsf{c}_2\frac{\d}{\d \textsf{c}_1})F(\textsf{c}_1,\textsf{c}_2)\;.$$
By stretching  variable $\textsf{c}_1 \rightarrow t\textsf{c}_1$, it is easy to see that the function
$$
\tilde{H}(\textsf{c}_1,\textsf{c}_2) = \int_0^1 dt \,t^{\alpha/\beta-1} \,\tilde{T}(t\textsf{c}_1,\textsf{c}_2)
$$
provides a solution to the last equation. By making a pullback map one solves the original
equation as
$$
H(\textsf{c}_1,\textsf{c}_2) = \frac{1}{\beta} \int_0^1 dt\,t^{\alpha/\beta-1}
\exp\Big(\frac{\gamma}{\beta}\,\frac{t-1}{t}\,\textsf{c}_2\frac{\d}{\d \textsf{c}_1}\Big)\,T(t\textsf{c}_1,\textsf{c}_2)\;.
$$
By substituting   particular values of parameters $\alpha = 2(d-3)$, $\beta=4$ and $\gamma=-4$, and
$T=T(\textsf{c}_1)$ one reproduces formula (\ref{dec_sol}).

\abz {\textsf{A.2.}} We use the following representations for the beta and gamma functions
\be
\nonumber
B(m,n) = \int_0^1 dt\, t^{m-1}(1-t)^{n-1}\;,\qquad {\rm Re}[m]>0, \;\;{\rm Re} [n]>0\;.
\ee
\be
\nonumber
B(m,n) = \frac{\Gamma(m)\Gamma(n)}{\Gamma(m+n)}\;.
\ee

\be
\nonumber
\ba{l}
\dps\Gamma(p/2+1)=\dps\sqrt{\pi}\dps\frac{p\,!!}{2^{(p+1)/2}}
\qquad (\textrm{odd} \quad p)\;,
\\
\\
\dps\Gamma(p/2+1)=\dps\frac{p\,!!}{2^{p/2}}
\qquad \quad\quad\quad (\textrm{even} \quad p)\;.
\ea
\ee

\vspace{5mm}



\begin{thebibliography}{99}
\addcontentsline{toc}{section}{References}

\parindent=0pt
\parskip=0pt



\bibitem{fronsdal}
C.~Fronsdal, Phys.\ Rev.  {\bf D18} (1978) 3624;
Phys.\ Rev.  {\bf D20} (1979) 848.


\bibitem{Curtright:1979uz}
  T.~Curtright,
  Phys.\ Lett.\  B {\bf 85} (1979) 219.



\bibitem{de Wit:1979pe}
  B.~de Wit and D.~Z.~Freedman,
  Phys.\ Rev.\  D {\bf 21} (1980) 358.



\bibitem{Siegel:1985tw}
  W.~Siegel and B.~Zwiebach,
  Nucl.\ Phys.\  B {\bf 263} (1986) 105;
  Nucl.\ Phys.\  B {\bf 282} (1987) 125.

\bibitem{Metsaev}
R.R. Metsaev, Phys.\ Lett. B \textbf{354} (1995) 78;
Phys.\ Lett.\ B {\bf 419} (1998) 49,
[arXiv:hep-th/9802097];
Talk given at International Seminar on Supersymmetries and Quantum
Symmetries (Dedicated to the Memory of Victor I. Ogievetsky),
Dubna, Russia, 22-26 Jul 1997,
[arXiv:hep-th/9810231].


\bibitem{Buchbinder:2001bs}
  I.~L.~Buchbinder, A.~Pashnev and M.~Tsulaia,
  Phys.\ Lett.\  B {\bf 523} (2001) 338
  [arXiv:hep-th/0109067].



\bibitem{Francia:2002aa}
  D.~Francia and A.~Sagnotti,
  Phys.\ Lett.\  B {\bf 543} (2002) 303
  [arXiv:hep-th/0207002].

\bibitem{Metsaev:2002vr}
  R.~R.~Metsaev,
  Phys.\ Lett.\  B {\bf 531} (2002) 152
  [arXiv:hep-th/0201226].

\bibitem{Vasiliev:1980as}
  M.~A.~Vasiliev,
  Yad.\ Fiz.\  {\bf 32} (1980) 855.

\bibitem{V1} M.A. Vasiliev, Fortsch.\ Phys.\ {\bf 35} (1987) 741.


\bibitem{Lopatin:1987hz}
  V.~E.~Lopatin and M.~A.~Vasiliev,
  Mod.\ Phys.\ Lett.\  A {\bf 3}, 257 (1988).


\bibitem{Vasiliev:2001wa}
  M.~A.~Vasiliev,
  Nucl.\ Phys.\ B {\bf 616} (2001) 106
  [Erratum-ibid.\ B {\bf 652} (2003) 407]
  [arXiv:hep-th/0106200].




\bibitem{Barnich:2004cr}
  G.~Barnich, M.~Grigoriev, A.~Semikhatov and I.~Tipunin,
  Commun.\ Math.\ Phys.\  {\bf 260} (2005) 147
  [arXiv:hep-th/0406192];
  G.~Barnich and M.~Grigoriev,
  JHEP {\bf 0608} (2006) 013
  [arXiv:hep-th/0602166].

\bibitem{Deser:1983mm}
  S.~Deser and R.~I.~Nepomechie,
  Annals Phys.\  {\bf 154} (1984) 396.


\bibitem{pmf}
   S.~Deser and A.~Waldron, Phys.\ Rev.\ Lett.\  {\bf 87} (2001) 031601
  [arXiv:hep-th/0102166]; Nucl.\ Phys.\  B {\bf 607}, 577 (2001)
  [arXiv:hep-th/0103198].

\bibitem{Hallowell:2005np}
  K.~Hallowell and A.~Waldron,
  Nucl.\ Phys.\  B {\bf 724} (2005) 453
  [arXiv:hep-th/0505255].


\bibitem{zin} Yu.~M.~Zinoviev, [arXiv:hep-th/0108192].

\bibitem{Skvortsov:2006at}
E.~D.~Skvortsov and M.~A.~Vasiliev, Nucl.\ Phys.\  B {\bf 756} (2006) 117
[arXiv:hep-th/0601095].


\bibitem{Konshtein:1988yg}
  S.~E.~Konshtein and M.~A.~Vasiliev,
  Nucl.\ Phys.\  B {\bf 312} (1989) 402.


\bibitem{Vasiliev:2003ev}
  M.~A.~Vasiliev,
  Phys.\ Lett.\ B {\bf 567} (2003) 139
  [arXiv:hep-th/0304049].


\bibitem{Vasiliev:2004cm}
  M.~A.~Vasiliev,
  JHEP {\bf 0412} (2004) 046
  [arXiv:hep-th/0404124].

\bibitem{Sagnotti:2005ns}
  A.~Sagnotti, E.~Sezgin and P.~Sundell,
  [arXiv:hep-th/0501156].

\bibitem{Eastwood:2002su}
  M.~G.~Eastwood,
  arXiv:hep-th/0206233.

\bibitem{Metsaev:2006ui}
  R.~R.~Metsaev,
  Phys.\ Rev.\  D {\bf 77} (2008) 025032
  [arXiv:hep-th/0612279].

\bibitem{Alkalaev:2002rq}
  K.~B.~Alkalaev and M.~A.~Vasiliev,
  Nucl.\ Phys.\ B {\bf 655} (2003) 57
  [arXiv:hep-th/0206068].

\bibitem{Alk_next}
  K.~B.~Alkalaev, in preparation.


\bibitem{ASV2}  K.B.~Alkalaev, O.V.~Shaynkman and M.A.~Vasiliev, JHEP \textbf{08} (2005) 069,
[arXiv:hep-th/0501108]; [arXiv:hep-th/0601225].


\bibitem{obzor}
X.~Bekaert, S.~Cnockaert, C.~Iazeolla, and M. A.~Vasiliev, "Nonlinear higher spin theories in various
dimensions, [arXiv:hep-th/0503128].


\bibitem{compensator} K. Stelle and P. West, Phys.\ Rev.\ D {\bf 21}
(1980) 1466; C.R. Preitschopf and M.A. Vasiliev, The Superalgebraic Approach to
Supergravity, in Proceedings of 31st International Ahrenshoop
Symposium On The Theory Of Elementary Particles, Berlin, Wiley-VCH,
1998, 496p  [arXiv: hep-th/9805127].


\bibitem{howe} R. Howe,  J. Amer. Math. Soc. 3 (1989) 2; Trans. Amer. Math. Soc. 2 (1989) 313.


\bibitem{Vasiliev:2005zu}
  M.~A.~Vasiliev,
  Int.\ J.\ Geom.\ Meth.\ Mod.\ Phys.\  {\bf 3} (2006) 37
  [arXiv:hep-th/0504090].


\bibitem{FradkinLin}
 E.~S.~Fradkin and V.~Y.~Linetsky,
  [Annals Phys.\  {\bf 198} (1990) 293].

\bibitem{Sezgin:2001zs}
  E.~Sezgin and P.~Sundell,
  JHEP {\bf 0109} (2001) 036
  [arXiv:hep-th/0105001].

\bibitem{Vasiliev:2001zy}
  M.~A.~Vasiliev,
  Phys.\ Rev.\  D {\bf 66} (2002) 066006
  [arXiv:hep-th/0106149].



\bibitem{Zwiebach:1992ie}
  B.~Zwiebach,
  Nucl.\ Phys.\  B {\bf 390} (1993) 33
  [arXiv:hep-th/9206084].



\bibitem{Buchbinder:2006eq}
  I.~L.~Buchbinder, A.~Fotopoulos, A.~C.~Petkou and M.~Tsulaia,
  Phys.\ Rev.\  D {\bf 74} (2006) 105018
  [arXiv:hep-th/0609082].

\bibitem{Fotopoulos:2007nm}
  A.~Fotopoulos and M.~Tsulaia,
  Phys.\ Rev.\  D {\bf 76} (2007) 025014
  [arXiv:0705.2939 [hep-th]].

\bibitem{Fotopoulos:2007yq}
  A.~Fotopoulos, N.~Irges, A.~C.~Petkou and M.~Tsulaia,
  JHEP {\bf 0710} (2007) 021
  [arXiv:0708.1399 [hep-th]].

\bibitem{Fotopoulos:2006ci}
  A.~Fotopoulos, K.~L.~Panigrahi and M.~Tsulaia,
  Phys.\ Rev.\  D {\bf 74} (2006) 085029
  [arXiv:hep-th/0607248].

\bibitem{Boulanger:2003vs}
  N.~Boulanger, S.~Cnockaert and M.~Henneaux,
  JHEP {\bf 0306} (2003) 060
  [arXiv:hep-th/0306023].



\end{thebibliography}
\end{document}